\RequirePackage[l2tabu, orthodox]{nag} %Checks dpreciated commands via l2tabu
\documentclass[10pt,letterpaper]{article}

\usepackage[plainpages=false,pdfpagelabels]{hyperref} % enables and customizes hyperlinks
\hypersetup{colorlinks = true, citecolor = blue, linkcolor = Maroon, urlcolor = Turquoise}

\usepackage{amsfonts,amssymb,amsmath,amsthm,mathtools} % AMS Math packages & Extension
\usepackage{mathrsfs}
\usepackage[dvipsnames]{xcolor} % enables colors
\usepackage[latin1]{inputenc} % Improve fonts used
\usepackage[T1]{fontenc}
\usepackage{lmodern}
\usepackage{graphicx}
\usepackage{epstopdf}
\usepackage{lipsum} %Lorem ipsum for testing
\usepackage{fixltx2e} %Fixes some LaTeX2e errors

\newtheorem{theorem}{Theorem}
\newtheorem{lemma}[theorem]{Lemma} % [theorem] ==> theorems and lemmas will share a counter

\newtheorem{corollary}[theorem]{Corollary}

\newtheorem{remark}[theorem]{Remark}
\newtheorem{assumption}[theorem]{Assumption}

\numberwithin{theorem}{section}
\numberwithin{equation}{section}

\usepackage[top=2.5cm, bottom=2.5cm, left=2.5cm, right=2.5cm]{geometry}
\usepackage{mathtools} % need for `show only references'
\mathtoolsset{showonlyrefs=true} % only equations which are labeled AND referenced will numbered % IMPORTANT NOTE...must use \eqref{} instead of (\ref{})

\usepackage{fixltx2e,amsmath} % Supposedly, this allows one to use \eqref{} in \caption{}.
\MakeRobust{\eqref}

\usepackage{dsfont} % gives you \mathds{} font %\allowdisplaybreaks % allows page breaks for long equations % you can prevent a page-break with \\*

\renewcommand{\[}{\left[}

\newcommand\Fb{\mathds{F}}

\newcommand\eps{\varepsilon}

\newcommand\sig{\sigma}

\def\ed{\;{\stackrel{\frak {D}}{=}}\;}
\title{Short-Time Expansions for Call Options on Leveraged ETFs under Exponential L\'{e}vy Models with Local Volatility}

\author{Jos\'{e} E. Figueroa-L\'{o}pez\thanks{Department of Mathematics, Washington University in St. Louis, St. Louis, MO 63130, USA.
{\tt figueroa@math.wustl.edu}. Research supported in part by the NSF Grants: DMS-1561141 and DMS-1613016.}
\and Ruoting Gong\thanks{Department of Applied Mathematics, Illinois Institute of Technology, Chicago, IL 60616, USA.
{\tt rgong2@iit.edu}.}
\and Matthew Lorig\thanks{Department of Applied Mathematics, University of Washington, Seattle, WA 98195, USA.
{\tt mlorig@uw.edu}.}}

\date{}

\begin{document}

\maketitle

\begin{abstract}
In this article, we consider the small-time asymptotics of options on a \emph{Leveraged Exchange-Traded Fund} (LETF) when the underlying Exchange Traded Fund (ETF) exhibits both local volatility and jumps of either finite or infinite activity. We show that leverage modifies the drift, volatility, jump intensity, and jump distribution of a LETF in addition to inducing the possibility of default, even when the underlying ETF price remains strictly positive. Our main results are closed-form expressions for the leading order terms of off-the-money European call and put LETF option prices near expiration, with explicit error bounds. These results show that the price of an out-of-the-money European call on a LETF with positive (negative) leverage is asymptotically equivalent, in short-time, to the price of an out-of-the-money European call (put) on the underlying ETF, but with modified spot and strike prices. Similar relationships hold for other off-the-money European options. These observations, in turn, suggest a method to hedge off-the-money LETF options near expiration using options on the underlying ETF. Finally, we derive a second order expansion for the implied volatility of an off-the-money LETF option and show both analytically and numerically how this is affected by leverage.
\end{abstract}

\vspace{0.3cm}


\vspace{0.3cm}


\section{Introduction}

A \emph{Leveraged Exchange-Traded Fund} (LETF) is a managed portfolio, which seeks to multiply the instantaneous returns of a reference Exchange-Traded Fund (ETF) by a constant \emph{leverage ratio} $\beta$. Typical values for $\beta$ are $\{-3,-2,-1,2,3\}$. The growing popularity of LETFs has led to the introduction of options written on these funds. As such, there has been much interest in how the leverage ratio $\beta$ affects both option prices and implied volatilities.

Cheng and Madhavan~\cite{ChengMadhavan:2009} and Avellaneda and Zhang~\cite{AvellanedaZhang:2010} are among the first to study LETFs. They notice that the terminal value of a LETF option depends not only on the terminal value of the underlying ETF, but also on the underlying ETF's integrated variance. Thus, European options on LETFs can be considered as \emph{path-dependent} options on the underlying ETF.

A variety of methods have been proposed for \emph{pricing} options on LETFs. Under the assumption that the ETF is a strictly positive diffusion with an independent volatility process, Zhu~\cite{Zhu:2007} shows that a European option on a LETF has the same price as a European option on an underlying ETF (with a different payoff function). Ahn, Haugh, and Jain~\cite{AhnHaughJain:2015} show that, when the underlying ETF has Heston dynamics, the corresponding LETF also has Heston dynamics, but with different parameters. Thus, in the Heston setting, options on LETFs can be priced using Fourier transforms. Additionally, they consider the case where the ETF follows Heston dynamics with independent compound Poisson jumps. They make the key insight that, when the underlying ETF can jump, the corresponding LETF could potentially jump to a negative value. As a result, the LETF manager must make payments to an insurer who guarantees that the value of the LETF portfolio never jumps to a negative value. An ad-hoc procedure for obtaining approximate option prices in this setting is also proposed in~\cite{AhnHaughJain:2015}.

There exists a number of studies that investigate how leverage affects \emph{implied volatility}. Under the assumption that the ETF is a diffusion process, Avellaneda and Zhang~\cite{AvellanedaZhang:2010} propose a formal scaling procedure to relate the implied volatilities of options on a LETF to those of options on its reference ETF. Leung and Sircar~\cite{LeungSircar:2015} study the implied volatility of LETF options assuming that the underlying ETF follows a fast mean-reverting volatility process, and obtain the same scaling as~\cite{AvellanedaZhang:2010}. Leung, Lorig, and Pascucci~\cite{LeungLorigPascucci:2016} study the implied volatility of LETF options assuming that the underlying ETF follows a general local-stochastic volatility model. They obtain the same scaling as~\cite{AvellanedaZhang:2010} at the zeroth-order, but caution that the scaling alone is not sufficient to capture the full effect of leverage on the implied volatility, and they derive higher order corrections to the scaling. Lee and Wang~\cite{LeeWang:2015} study how leverage affects implied volatility, and relate the implied volatility surfaces of the leveraged product and the underlying, in different asymptotic regimes, via shifting/scaling transforms. The models considered by~\cite{LeeWang:2015} include stochastic volatility models, models with fractional Brownian motion volatility, and exponential L\'{e}vy models. For exponential L\'{e}vy models, Lee and Wang~\cite{LeeWang:2015} assume that the support of the L\'{e}vy measure is bounded below (respectively, above) in the case of positive (respectively, negative) leverage ratio, so that jumps in the underlying ETF never cause the corresponding LETF to jump to zero or a negative value (and thus, avoid the insurance payments considered in~\cite{AhnHaughJain:2015}). We mention, finally, that there is a recent book~\cite{LeungSantoli:2016} by Leung and Santoli examining various aspects of LETFs.

In this article, we consider the small-time asymptotics of LETF options when the underlying ETF exhibits both local volatility and jumps of either
finite or infinite activity. Besides being a mathematically challenging framework to work with, local volatility models with L\'evy type jumps offer several benefits over purely local volatility models such as increased stability of the calibrated local coefficient through time and a better fit of the steep volatility smiles observed at short maturities (see, for instance, \cite{Andersen2000}). It is worth mentioning here that, as with a purely local volatility model, given a parametrically specified jump component, it is possible to formally deduce a volatility coefficient that can perfectly calibrate an observed smoothly interpolated implied volatility smile for a fixed maturity. This can be done via an analogous Dupire formula (cf.~\cite{Dupire}) for local volatility models with jumps (see \cite[Proposition 3]{ContBentata:2015} for details).

In the context of local volatility models with jumps, the leverage ratio creates the following effects in the risk-neutral dynamics of the LETF. Firstly, the leverage ratio induces in the LETF the possibility of default even when the underlying ETF cannot default, which in turn modifies the risk-neutral drift of the LETF. Moreover, the leverage ratio modifies both the distribution and the intensity of jumps of the LETF. In particular, the L\'{e}vy density of the LETF may not have a full support and may not be smooth, even when the L\'{e}vy density of the underlying ETF has a full support on $\mathbb{R}$ and is smooth. Our analysis shows rigorously how the above effects transform option prices and implied volatilities. Our main results provide closed-form expressions of the leading-order term of off-the-money European call and put LETF option prices, near expiration. As in a local jump-diffusion model (cf.~\cite{FigueroaLopezLuoOuyang:2014}), we show that the option prices are asymptotically equivalent to $b_{1}t$, with $t$ representing the time-to-maturity and $b_{1}$ being a specified constant, which only depends on the jump component of the process. Precise error bounds are also provided. On one hand, our results uncover a puzzling and useful connection, near expiration, between the prices of options written on a LETF and those of options written on the underlying ETF. On the other hand, the results therein are accurate enough to enable us to find a close-to-maturity expansion for the implied volatility of ``arbitrary order'' along the lines of Gao and Lee~\cite{GaoLee:2014}. For simplicity and completeness, in this work we derive the second-order expansion for the implied volatility, which, as expected, is similar to that of an exponential L\'{e}vy model (cf.~\cite{FigueroaLopezForde:2012}) and sheds some light on the behavior of the implied volatility surface for LETF options near expiration. In particular, we show that the leverage coefficient only appears in the second-order term and we explicitly illustrate how the leverage coefficient affects the behavior of this term.

Let us briefly comment on the connection of our work with some related literature and highlight some technical difficulties specific to our work.  Like~\cite{AhnHaughJain:2015}, we allow for the possibility that, in the absence of insurance, a jump in the ETF could cause the corresponding LETF to jump to a negative value. Thus, our results are fundamentally different from those of~\cite{LeeWang:2015}, who do not allow for this possibility and are the only authors that study the short-time asymptotics of the implied volatility of LETFs in a jump setting. Let us remark that the local volatility framework adopted in our work does not allow us to use the built-in expansions of other frameworks studied before, because, when the underlying ETF exhibits local volatility, the resulting LETF option prices cannot be framed as options on a single asset following its own Markovian dynamics. By contrast, when the underlying ETF exhibits local volatility, the LETF option prices resemble those of options on a stochastic volatility process $Y$ with jumps, in which the volatility is driven by another process $X$, whose Brownian and jump components are perfectly correlated with those of the underlying asset $Y$. To the best of our knowledge, this framework has not been considered in the literature of short-time asymptotics. In particular, due to the perfect correlation of the noise and jumps as well as the singularity of the jump coefficient of the LETF, there is limited information about the transition densities of $(X,Y)$ that is available, starting with its existence and, moreover, its required regularity that was used in earlier works such as in~\cite{GatheralHsuLaurenceOuyangWang:2012} and~\cite{FigueroaLopezLuoOuyang:2014}. To overcome this difficulty, we approximate the option prices, up to a $O(t^{3/2})$ term, by the price of a simple European claim on $(X,Y)$ with a sufficiently smooth payoff function.

The rest of this paper proceeds as follows. In Section \ref{sec:PrelimResult} we set up the LETF option pricing problem, establish some notation and provide some preliminary results, which shall be needed in subsequent sections. In Section \ref{sec:OptLETFs} we derive explicit small-time expansions for off-the-money LETF option prices and provide asymptotic error bounds for these expansions (see Theorem \ref{thm:SmallTimeOTMCallOptLETF}).
In Section \ref{sec:IVLETFs} we translate our small-time option price expansion into a small-time expansion of implied volatility. Finally, in Section \ref{sec:numerical} we implement our implied volatility expansion in two numerical examples.

\section{Setup and Preliminary Results}\label{sec:PrelimResult}

Throughout this article, $C^{n}(\mathbb{R})$, $n\in\mathbb{Z}^{+}:=\mathbb{N}\cup\{0\}$, is the class of real-valued functions, defined on $\mathbb{R}$, which have continuous derivatives of order $k=0,\ldots,n$, while $C_{b}^{n}(\mathbb{R})\subset C^{n}(\mathbb{R})$ corresponds to those functions having bounded derivatives. In a similar fashion, $C^{\infty}(\mathbb{R})$ is the class of real-valued functions, defined on $\mathbb{R}$, which have continuous derivatives of any order $k\in\mathbb{Z}^{+}$, while $C_{b}^{\infty}(\mathbb{R})\subset C^{\infty}(\mathbb{R})$ are again the functions having bounded derivatives. Sometimes, $\mathbb{R}$ will be replaced by $\mathbb{R}_{0}:=\mathbb{R}\setminus\{0\}$ or $\mathbb{R}^{n}$, when the functions are defined on these spaces.

\subsection{The Dynamics of Leveraged ETFs}\label{subsec:DynLETFs}

Throughout this paper, {let} $(\Omega,\mathscr{F},\mathbb{F},\mathbb{P})$ be a complete filtered probability space. The filtration
$\mathbb{F}:=(\mathscr{F}_{t})_{t\geq 0}$ represents the history of the market. All stochastic processes defined below live on this probability space, and, unless otherwise indicated, all expectations are taken with respect to $\mathbb{P}$, where $\mathbb{P}$ represents the risk-neutral probability measure of the market. For simplicity, we assume a frictionless market, no arbitrage, zero interest rates and no dividends.

Let $W:=(W_{t})_{t\geq 0}$ be a standard Brownian motion with respect to $\mathbb{F}$ under $\mathbb{P}$. Let $N(dt,dz)$ be a Poisson random measure
on $[0,\infty)\times\mathbb{R}_{0}$ under $\mathbb{P}$ with mean measure $dt\,\nu(dz)$, where $\nu$ is a L\'{e}vy measure (i.e., $\nu$ is such
that $\int_{\mathbb{R}_{0}}(|x|^{2}\wedge 1)\nu(dx)<\infty$). The compensated Poisson random measure of $N$ is denoted by $\widetilde{N}$. Assume that $W$ and $N$ are independent under $\mathbb{P}$. Without loss of generality, we also assume that $N$ is the jump measure of a L\'{e}vy process $Z:=(Z_{t})_{t\geq 0}$ with L\'{e}vy measure $\nu$.

Consider an Exchange-Traded Fund (ETF), whose price process $S:=(S_{t})_{t\geq 0}$ has dynamics, under the pricing measure $\mathbb{P}$, of the form
\begin{align}\label{eq:dS}
\text{ETF}:\qquad\qquad dS_{t}=S_{t-}\left(\sigma_{t}\,dW_{t}+\int_{\mathbb{R}_{0}}\left(e^{z}-1\right)\widetilde{N}(dt,dz)\right),
\end{align}
where we are implicitly assuming that $\nu$ satisfies the integrability condition
\begin{equation}\label{FAONNu}
\int_{|z|>1}e^{z}\nu(dz)<\infty.
\end{equation}
We shall impose below further assumptions on $\nu$ and $\sigma:=(\sigma_{t})_{t\geq 0}$ (see Assumption \ref{assump:UnifBdVol} and Assumption \ref{assump:SmthBdLevyDen}) so that $S$ is a true $\mathbb{F}$-martingale under $\mathbb{P}$. Let $L:=(L_{t})_{t\geq 0}$ be the price process of a Leveraged Exchange-Traded Fund (LETF) with underlying $S$ and leverage ratio $\beta\in\mathbb{R}$. Typical values of $\beta$ are $\{-3,-2,-1,2,3\}$. Throughout this article, we assume that
\begin{align}
\beta\in(-\infty,-1]\cup[1,\infty),
\end{align}
as no LETFs are traded with leverage $\beta\in(-1,1)$. Concretely, the dynamics of $L$ under $\mathbb{P}$ are as follows:
\begin{align}\label{eq:dL}
\text{LETF}:\qquad dL_{t}=\beta\frac{L_{t-}}{S_{t-}}\,dS_{t}+L_{t-}\,d\widetilde{M}_{t},\qquad
d\widetilde{M}_{t}=-\int_{A^{c}}\left[\beta\left(e^{z}-1\right)+1\right]\widetilde{N}(dt,dz),
\end{align}
where
\begin{align}
A:=\left\{z\in\mathbb{R}:\,\beta\left(e^{z}-1\right)+1>0\right\},\qquad A^{c}:=\left\{z\in\mathbb{R}:\,\beta\left(e^{z}-1\right)+1\leq 0\right\}.
\end{align}
Let us explain the intuition behind the dynamics of $L$. A LETF manager seeks to provide investors with a portfolio that multiplies the instantaneous
returns of $S$ by the leverage ratio $\beta$. To do this, at time $t$, the manager holds $\Delta_{t}=\beta(L_{t-}/S_{t-})$ shares of the ETF $S$. Thus, the change in the value of $L$ due to changes in the value of $S$ is $\beta(L_{t-}/S_{t-})dS_{t}$, which explains the first term in \eqref{eq:dL}. For the second term in \eqref{eq:dL}, note that, in the absence of such a term, we would have that
\begin{align}
L_{t}=L_{t-}+\Delta L_{t}
=L_{t-}+\beta\frac{L_{t-}}{S_{t-}}\Delta S_{t}=L_{t-}+ L_{t-} \beta\left(e^{\Delta Z_{t}}-1\right)=L_{t-}\left[\beta\left(e^{\Delta Z_{t}}-1\right)+1\right].
\end{align}
The last quantity above would then be zero or negative if $\Delta Z_{t}\in A^{c}$. In order to prevent $L$ from becoming negative, the LETF manager must make continuous payments at a certain rate $\lambda_{t}$ to an insurer who, in the event that $\Delta Z_{t}\in A^{c}$, must pay $-L_{t-}[\beta(e^{\Delta Z_{t}}-1)+1]$ to the LETF manager so that the portfolio value becomes exactly zero. The payments $\lambda_{t}dt$ made by the LETF manager to the insurer in the interval $[t,t+dt)$ must be equal to the expected amount paid by the insurer in this interval under $\mathbb{P}$, i.e.,
\begin{align}
\lambda_{t}\,dt=\mathbb{E}\left(\left.-L_{t-}\int_{A^{c}}\left[\beta\left(e^{z}-1\right)+1\right]N(dt,dz)\,\right|L_{t-}\right)=-L_{t-}\int_{A^{c}}\left[\beta\left(e^{z}-1\right)+1\right]\nu(dz)\,dt.
\end{align}
Thus, the net cash flow from the LETF manager to the insurer in the interval $[t,t+dt)$ is
\begin{align}
-\lambda_{t}\,dt-L_{t-}\int_{A^{c}}\left[\beta\left(e^{z}-1\right)+1\right]N(dt,dz)=-L_{t-}\int_{A^{c}}\left[\beta\left(e^{z}-1\right)+1\right]\widetilde{N}(dt,dz)=L_{t-}\,d\widetilde{M}_{t},
\end{align}
where, in the last equality, we have used the definition of $\widetilde{M}$ as given in \eqref{eq:dL}. Combining the cash flow from the LETF manager
to the insurer with the levered position in the ETF, we obtain the dynamics \eqref{eq:dL} for $L$.

It will be helpful to have a more explicit expression for the dynamics of $L$. Plugging the expressions for $dS_{t}$ and $d\widetilde{M}_{t}$ into the expression for $dL_{t}$ in \eqref{eq:dL}, we obtain that
\begin{align}\label{eq:dL.2}
dL_{t}=L_{t-}\left[\beta\sigma_{t}\,dW_{t}+\int_{A_{0}}\beta\left(e^{z}-1\right)\widetilde{N}(dt,dz)-\int_{A^{c}}\widetilde{N}(dt,dz)\right],
\end{align}
where hereafter $A_{0}:=A\backslash\{0\}$. From \eqref{eq:dL.2}, we observe that $L$ jumps to zero exactly when $\Delta Z_{t}\in A^{c}$. Thus, we
define the \emph{default time} of $L$ as
\begin{align}\label{eq:DefaultTimeLETF}
\tau:=\inf\left\{t\geq 0:\,\Delta Z_{t}\in A^{c}\right\}.
\end{align}
Note that the \emph{default intensity} of $L$ is $\nu(A^c)$, which is finite since $A^{c}\cap[-\eps,\eps]=\emptyset$ for some $\eps>0$ small enough. By a simple application of It\^{o}'s Lemma and assuming for simplicity that $\int_{|z|\geq{}1}|z|\nu(dz)<\infty$, the dynamics of $S$ and $L$ can respectively be written as
\begin{align}
\text{ETF}:&& S_{t}&=e^{X_{t}(x)}, &
X_{t}(x)&=x+\int_{0}^{t}\mu_{s}\,ds+\int_{0}^{t}\sigma_{s}\,dW_{s}+\int_{0}^{t}\int_{\mathbb{R}_{0}}z\,\widetilde{N}(ds,dz),\\
\text{LETF}:&& L_{t}&={\bf 1}_{\{\tau>t\}}e^{Y_{t}(x)}, &
Y_{t}(x)&=x+\int_{0}^{t}\gamma_{s}\,ds+\beta\int_{0}^{t}\sigma_{s}\,dW_{s}+\int_{0}^{t}\int_{A_{0}}\ln\left(\beta\left(e^{z}-1\right)+1\right)\widetilde{N}(ds,dz),
\end{align}
where the drifts $\mu_{t}$ and $\gamma_{t}$ are given by
\begin{align}\label{eq:MuGamma}
\mu_{t}:=-\frac{1}{2}\sigma_{t}^{2}-\!\int_{\mathbb{R}_{0}}\!\left(e^{z}-1-z\right)\nu(dz),\quad\gamma_{t}:=\nu(A^{c})-\frac{1}{2}\beta^{2}\sigma_{t}^{2}-\!\int_{A_{0}}\!\left[\beta\left(e^{z}-1\right)-\ln\left(\beta\left(e^{z}-1\right)+1\right)\right]\nu(dz).
\end{align}
In what follows, we will refer to $X(x):=(X_{t}(x))_{t\geq 0}$ and $Y(x):=(Y_{t}(x))_{t\geq 0}$ as the ``log-ETF'' process and the ``log-LETF'' process, respectively. For convenience, we will omit the variable $x$ if there is no risk of confusion. Moreover, we will sometimes use the phrase ``option on $X$'' to mean ``option on $S$'', and likewise for $L$ and $Y$.
\begin{remark}\label{rem:JumpSupport}
For any fixed $\beta\in(-\infty,-1]\cup[1,\infty)$, define
\begin{align}\label{eq:ubetaz}
u_{\beta}(z):=\ln\left(\beta\left(e^{z}-1\right)+1\right),\quad z\in A.
\end{align}
Note that, when $X$ experiences a jump of size $z\in A$, $Y$ experiences a jump of size $u_{\beta}(z)$. It follows that
\begin{align}
\beta\geq 1:&& A&=\left(\ln\left(1-\beta^{-1}\right),\infty\right), & u_{\beta}(A)&=\mathbb{R}\backslash\{0\},\\
\beta\leq -1:&& A&=\left(-\infty,\ln\left(1-\beta^{-1}\right)\right), & u_{\beta}(A)&=\left(-\infty,\ln(1-\beta)\right).
\end{align}
In particular, when $\beta\leq -1$, the jumps of the process $Y$ are limited to sizes $z<\ln(1-\beta)$.
\end{remark}

Note that, if the volatility process $\sigma$ were constant, then both $X$ and $Y$ would be L\'{e}vy processes with respective L\'{e}vy triplets
$(\mu,\sigma^{2},\nu)$ and $(\gamma,\beta^{2}\sigma^{2},\nu\circ u_{\beta}^{-1})$. In this case, options on $X$ and options on $Y$ could be analyzed
independently using standard theory. However, as has been widely documented in the literature, it is not realistic to assume that the volatility process $\sigma$ is constant, as this would result in options prices that are inconsistent with the observed term-structure of implied volatility. Of particular relevance are \emph{local volatility} dynamics, which are known to be able to perfectly replicate the implied volatility surface at any given time. With this in mind, we hereafter adopt the following setup:
\begin{assumption}\label{assump:UnifBdVol}
The volatility process $\sigma$ is of the form $\sigma_{t}=\sigma(X_{t})$, for any $t\geq 0$, where $\sigma(\cdot)\in C_{b}^{\infty}(\mathbb{R})$ is a deterministic function.
\end{assumption}
The dynamics of the ETF and the LETF can then be written as
\begin{align}
\label{eq:DynETF} \text{ETF}:\quad
&S_{t}=e^{X_{t}(x)},\qquad\quad\,\,\,X_{t}(x)=x+\int_{0}^{t}\mu(X_{s})\,ds+\int_{0}^{t}\sigma(X_{s})\,dW_{s}+\int_{0}^{t}\int_{\mathbb{R}_{0}}z\,\widetilde{N}(ds,dz),\\
\label{eq:DynLETF} \text{LETF}:\quad &L_{t}={\bf 1}_{\{\tau>t\}}e^{Y_{t}(x)},\quad
Y_{t}(x)=x+\int_{0}^{t}\gamma(X_{s})\,ds+\beta\!\int_{0}^{t}\sigma(X_{s})\,dW_{s}+\int_{0}^{t}\!\int_{A_{0}}\!u_{\beta}(z)\,\widetilde{N}(ds,dz),
\end{align}
for $t\geq 0$, where
\begin{align}
\mu(u):=-\frac{1}{2}\sigma^{2}(u)-\int_{\mathbb{R}_{0}}\left(e^{z}-1-z\right)\nu(dz),\quad\gamma(u)&:=\nu(A^{c})-\frac{1}{2}\beta^{2}\sigma^{2}(u)-\int_{A_{0}}\left[\beta\left(e^{z}-1\right)-u_{\beta}(z)\right]\nu(dz),
\end{align}
and where $u_{\beta}(z)$ is defined in \eqref{eq:ubetaz}. Compared with their constant volatility counterparts, local volatility models are able to better capture the term-structure of implied volatility. Under local volatility dynamics, the process $Y$ alone is \emph{not} a Markov process but the pair $(X,Y)$ is. Thus, to analyze options on $Y$ we must consider the pair $(X,Y)$ jointly.
\begin{remark}\label{rem:ExistUniqSDESol}
Assumption \ref{assump:UnifBdVol} guarantees that the SDEs \eqref{eq:DynETF} and \eqref{eq:DynLETF} admit a unique strong solution (cf.~\cite[Theorem
6.2.3]{Applebaum:2009} and~\cite[Theorem 1.19]{OksendalSulem:2005}).
\end{remark}

We will also impose the following conditions on the L\'{e}vy measure, which collect and extend some of the conditions mentioned above.
\begin{assumption}\label{assump:SmthBdLevyDen}
The L\'{e}vy measure $\nu$ admits a $C^{2}(\mathbb{R}_{0})$ density $h$, i.e., $\nu(dz)=h(z)dz$. Moreover, the L\'{e}vy density $h$ satisfies the following conditions:
\begin{align}
&{\rm (i)}\,\int_{{\{|z|>1\}}}{|z|}h(z)\,dz<\infty;\\
&{\rm (ii)}\,\int_{{\{z>1\}}}{e^{(1+\delta)z}}h(z)\,dz<\infty,\;\text{ for some }\delta>0;\\
&{\rm (iii)}\,\sup_{|z|>\varepsilon}\left|h^{(n)}(z)\right|<\infty,\;\text{ for any }\,\varepsilon>0\text{ and }\,n=0,1,2.
\end{align}
\end{assumption}
\begin{remark}\label{rem:SmthBdLevyDen}
Assumption \ref{assump:SmthBdLevyDen}-(ii) is only needed for the case of $\beta\geq 1$ to prove Lemma \ref{lem:SmallTimeEstI3} below. This condition slightly strengthens the well-known condition $\int_{\{z>1\}}e^{z}h(z)dz<\infty$, which is needed for $S_{t}=e^{X_{t}}$ to have a finite mean. Assumption \ref{assump:SmthBdLevyDen}-(iii) is crucial for the tail probabilities $\mathbb{P}(Y_{t}\geq y)$, $y>0$, of $Y$ to vanish to $0$ at the order of $O(t)$, as $t\rightarrow 0$. Indeed, even in the simplest pure-jump L\'{e}vy case, it is possible to build examples where the tail probability converges to $0$, as $t\rightarrow 0$, as a fraction power of $t$ in the absence of Assumption \ref{assump:SmthBdLevyDen}-(iii) (cf.~\cite{Marchal:2009}).
\end{remark}

\subsection{Notations}\label{subsec:Notations}

In this subsection, we introduce the definitions of some important processes. For any $x\in\mathbb{R}$, let $\widetilde{X}(x):=(\widetilde{X}_{t}(x))_{t\geq 0}$ and $\widetilde{Y}(x):=(\widetilde{Y}_{t}(x))_{t\geq 0}$ be the solution of the following two-dimensional SDE
\begin{align}\label{eq:DynTildeY}
\widetilde{Y}_{t}(x)&=x+\int_{0}^{t}\gamma\!\left(\widetilde{X}_{s}(x)\right)ds+\beta\int_{0}^{t}\sigma\!\left(\widetilde{X}_{s}(x)\right)dW_{s}+\int_{0}^{t}\int_{A_{0}}u_{\beta}(z)\,\widetilde{N}(ds,dz),
& t &\geq 0,\\
\widetilde{X}_{t}(x)&=x+\int_{0}^{t}\widetilde{\mu}\!\left(\widetilde{X}_{s}(x)\right)ds+\int_{0}^{t}\sigma\!\left(\widetilde{X}_{s}(x)\right)dW_{s}+\int_{0}^{t}\int_{A_{0}}z\,\widetilde{N}(ds,dz),
& t &\geq 0,
\end{align}
where
\begin{align}
\widetilde{\mu}(u):=\mu(u)-\int_{A^{c}}z\,\nu(dz)=-\frac{1}{2}\sigma^{2}(u)-\int_{\mathbb{R}_{0}}\left(e^{z}-1-z\right)\nu(dz)-\int_{A^{c}}z\,\nu(dz).
\end{align}
Note that, for any $t\geq 0$, we have
\begin{align}
\left(\widetilde{X}_{s}(x),\widetilde{Y}_{s}(x)\right)_{s\in[0,t]}\ed\left(\left.\left(X_{s}(x),Y_{s}(x)\right)_{s\in[0,t]}\,\right|N([0,t]\times{}A^{c})=0\right).
\end{align}
Moreover, the pair $(\widetilde{X},\widetilde{Y})$ can be seen as a stochastic volatility model, where the driver of the volatility, $\widetilde{X}$,
has a L\'{e}vy jump component. Note that the jump and continuous components of the processes $\widetilde{Y}$ and $\widetilde{X}$ are correlated with
each other.

Let
\begin{align}
Z_{t}(A):=\int_{0}^{t}\int_{A_{0}}z\widetilde{N}(ds,dz),\quad t\geq 0,
\end{align}
be the underlying L\'evy process driving the dynamics of the processes $(\widetilde{X},\widetilde{Y})$. As is usually the case, we will decompose
$Z(A):=(Z_{t}(A))_{t\geq 0}$ into a compound Poisson process and a process with bounded jumps (cf.~\cite{Leandre:1987} and~\cite{FigueroaLopezLuoOuyang:2014}). More precisely, for any $\varepsilon\in(0,|\ln(1-1/\beta)|\wedge 1)$, let $c_{\varepsilon}\in
C^{\infty}(\mathbb{R})$ be a truncation function such that ${\bf 1}_{[-\varepsilon/2,\varepsilon/2]}\leq c_{\varepsilon}\leq {\bf
1}_{[-\varepsilon,\varepsilon]}$. Let $Z^{\varepsilon,1}(A):=(Z^{\varepsilon,1}_{t}(A))_{t\geq 0}$ and $Z^{\varepsilon,2}(A):=Z^{\varepsilon,2}_{t}(A))_{t\geq 0}$ be two independent L\'{e}vy processes with respective L\'{e}vy triplets $(b_{\varepsilon},0,\nu_{A}^{\varepsilon,1}(dz))$ and $(0,0,\nu_{A}^{\varepsilon,2}(dz))$, where
\begin{align}
\nu_{A}^{\varepsilon,1}(dz):=h_{A}^{\varepsilon,1}(z)\,dz:={\bf 1}_{A}(z)c_{\varepsilon}(z)h(z)\,dz,\quad\nu_{A}^{\varepsilon,2}(dz):=h_{A}^{\varepsilon,2}(z)\,dz:={\bf
1}_{A}(z)\left(1-c_{\varepsilon}(z)\right)h(z)\,dz,
\end{align}
and
\begin{align}
b_{\varepsilon}:=-\int_{A\setminus[-1,1]}zh(z)\,dz-\int_{A\cap[-1,1]}z\left(1-c_{\varepsilon}(z)\right)h(z)\,dz.
\end{align}
Moreover, let $Z^{\varepsilon}(A):=(Z^{\varepsilon}_{t}(A))_{t\geq 0}$ be the process defined by
\begin{align}\label{eq:DecompZA}
Z^{\varepsilon}_{t}(A):=Z^{\varepsilon,1}_{t}(A)+Z^{\varepsilon,2}_{t}(A),\quad t\geq 0.
\end{align}
Clearly, $Z(A)$ has the same law as $Z^{\varepsilon}(A)$. The process $Z^{\varepsilon,1}(A)$, which hereafter we refer to as the \emph{small-jump component} of $Z(A)$, is a pure-jump L\'{e}vy process with jumps bounded by $\varepsilon$. By contrast, the process $Z^{\varepsilon,2}(A)$, hereafter referred to as the \emph{big-jump component} of $Z(A)$, is a compound Poisson process with intensity of jumps $\nu_{A}^{\varepsilon,2}(A)$, and jumps $(J^{(i)}_{\varepsilon})_{i\geq 1}$ with probability density function
\begin{align}\label{eq:DenJumpZ2epsA}
g_{J}(z;\varepsilon,\beta):=\frac{h_{A}^{\varepsilon,2}(z)}{\nu_{A}^{\varepsilon,2}(A)}=\frac{1}{\nu_{A}^{\varepsilon,2}(A)}{\bf
1}_{A}(z)\left(1-c_{\varepsilon}(z)\right)h(z).
\end{align}
Throughout this paper, we denote by $N^{\varepsilon}(A):=(N_{t}^{\varepsilon}(A))_{t\geq 0}$ and $\lambda_{\varepsilon}(A):=\nu_{A}^{\varepsilon,2}(A)$, respectively, the jump counting process and the jump intensity of the compound Poisson
process $Z^{\varepsilon,2}(A)$, and by $(\tau_{i})_{i\geq 1}$ the jump times of $Z^{\varepsilon,2}(A)$.

Let $M_{A}^{\varepsilon}$ and $M^{\varepsilon,1}_{A}$ denote the respective jump measures of $Z^{\varepsilon}(A)$ and $Z^{\varepsilon,1}(A)$, and let
$\widetilde{M}_{A}^{\varepsilon}$ and $\widetilde{M}_{A}^{\varepsilon,1}$ be the respective compensated random measures. Let
$\widetilde{W}:=(\widetilde{W}_{t})_{t\geq 0}$ be a standard Brownian motion independent of $W$. Consider the processes $Y^{o}(x,y):=(Y_{t}^{o}(x,y))_{t\geq 0}$ and $X^{o}(x):=(X_{t}^{o}(x))_{t\geq 0}$ defined as the solution of the two-dimensional SDE
\begin{align}\label{eq:SDEYo}
Y_{t}^{o}(x,y)&=y+\int_{0}^{t}\gamma(X^{o}_{s}(x))\,ds+\beta\int_{0}^{t}\sigma(X^{o}_{s}(x))\,d\widetilde{W}_{s}+\int_{0}^{t}\int_{\mathbb{R}_{0}}u_{\beta}(z)\,\widetilde{M}_{A}^{\varepsilon}(ds,dz),
& t &\geq 0,\\
\label{eq:SDEXo}X_{t}^{o}(x)&=x+\int_{0}^{t}\widetilde{\mu}(X^{o}_{s}(x))\,ds+\int_{0}^{t}\sigma(X^{o}_{s}(x))\,d\widetilde{W}_{s}+\int_{0}^{t}\int_{\mathbb{R}_{0}}z\,\widetilde{M}_{A}^{\varepsilon}(ds,dz),
& t &\geq 0.
\end{align}
Since $Z(A)$ has the same law as $Z^{\varepsilon}(A)$, it follows that $(X^{o}(x),Y^{o}(x,x))$ has the same law as $(\widetilde{X}(x),\widetilde{Y}(x))$. Hence, in order to study the small-time asymptotics of an option on $\widetilde{Y}(x)$, we can (and will) analyze the behavior of the same option on $Y^{o}(x,x)$.

Next, for any fixed $\varepsilon>0$, define the processes $Y^{\varepsilon}(x,y):=(Y_{t}^{\varepsilon}(x,y))_{t\geq 0}$ and
$X^{\varepsilon}(x):=(X_{t}^{\varepsilon}(x))_{t\geq 0}$ as the solution of the two-dimensional SDE
\begin{align}\label{eq:SDEYeps}
Y_{t}^{\varepsilon}(x,y)&=y+\int_{0}^{t}\gamma_{\varepsilon}(X^{\varepsilon}_{s}(x))\,ds+\beta\int_{0}^{t}\sigma(X^{\varepsilon}_{s}(x))\,d\widetilde{W}_{s}+\int_{0}^{t}\int_{\mathbb{R}_{0}}u_{\beta}(z)\,\widetilde{M}^{\varepsilon,1}_{A}(ds,dz),\\
\label{eq:SDEXeps} X^{\varepsilon}_{t}(x)&=x+\int_{0}^{t}\mu_{\varepsilon}(X^{\varepsilon}_{s}(x))\,ds+\int_{0}^{t}\sigma(X^{\varepsilon}_{s}(x))\,d\widetilde{W}_{s}+\int_{0}^{t}\int_{\mathbb{R}_{0}}z\,\widetilde{M}^{\varepsilon,1}_{A}(ds,dz),
\end{align}
where
\begin{align}\label{Eq:Dfngammaepsilon}
\gamma_{\varepsilon}(u)&:=\gamma(u)-\int_{A}u_{\beta}(z)\left(1-c_{\varepsilon}(z)\right)h(z)\,dz\\
&\,\,=\nu(A^{c})-\frac{1}{2}\beta^{2}\sigma^{2}(u)-\int_{A_{0}}\left[\beta\left(e^{z}-1\right)-u_{\beta}(z)\right]h(z)\,dz-\!\int_{A}u_{\beta}(z)\left(1-c_{\varepsilon}(z)\right)h(z)\,dz,\\
\mu_{\varepsilon}(u)&:=\widetilde{\mu}(u)-\int_{A}z\left(1-c_{\varepsilon}(z)\right)h(z)\,dz\\
&\,\,=-\frac{1}{2}\sigma^{2}(u)-\int_{\mathbb{R}_{0}}\left(e^{z}-1-z\right)h(z)\,dz-\int_{A}z\left(1-c_{\varepsilon}(z)\right)h(z)\,dz-\int_{A^{c}}zh(z)\,dz.
\end{align}
As observed from \eqref{eq:SDEYo} and \eqref{eq:SDEXo}, the law of the processes \eqref{eq:SDEYeps} and \eqref{eq:SDEXeps} up to time $t$ can be
interpreted as the law of $(X_{s}^{o}(x),Y_{s}^{o}(x,y))_{s\in[0,t]}$ conditioned on not having any ``big'' jumps in $[0,t]$. In other words,
for any $t\geq 0$, we have
\begin{align}
\left(X^{\varepsilon}_{s}(x),Y_{s}^{\varepsilon}(x,y)\right)_{s\in[0,t]}\ed\left(\left.\left(X_{s}^{o}(x),Y_{s}^{o}(x,y)\right)_{s\in[0,t]}\,\right|N^{\varepsilon}_{t}(A)=0\right).
\end{align}
The processes defined above will be needed when we expand the moments of $Y_{t}^{o}(x,x)$ in powers of time by conditioning on the number of jumps of $Z^{\varepsilon,2}(A)$.

\subsection{The Dynkin's Formula}\label{subsec:Dynkin}

For future reference, we now proceed to state a Dynkin's formula for the ``small-jump'' pair $(X^{\varepsilon}(x),Y^{\varepsilon}(x,y))$, defined in
\eqref{eq:SDEYeps}-\eqref{eq:SDEXeps}. To this end, let us first remark that the infinitesimal generator of $(X^{\varepsilon}(x),Y^{\varepsilon}(x,y))$, hereafter denoted by $L_{\varepsilon}$, can be written as:
\begin{align}\label{eq:Leps}
L_{\varepsilon}f(x,y)=\mathcal{D}_{\varepsilon}f(x,y)+\mathcal{I}_{\varepsilon}f(x,y),\quad f\in C_{b}^{2}(\mathbb{R}^{2}),
\end{align}
where
\begin{align}\label{eq:Deps}
\mathcal{D}_{\varepsilon}f(x,y)&:=\mu_{\varepsilon}(x)\!\frac{\partial f}{\partial x}(x,\!y)\!+\!\gamma_{\varepsilon}(x)\!\frac{\partial f}{\partial
y}(x,\!y)\!+\!\frac{\sigma^{2}\!(x)}{2}\!\frac{\partial^{2}f}{\partial x^{2}}(x,\!y)\!+\!\frac{\beta^{2}\!\sigma^{2}\!(x)}{2}\!\frac{\partial^{2}f}{\partial y^{2}}(x,\!y)\!+\!\beta\sigma^{2}\!(x)\!\frac{\partial^{2}f}{\partial x\partial y}(x,\!y),\\
\label{eq:Ieps} \mathcal{I}_{\varepsilon}f(x,y)&:=\int_{A_{0}}\left[f\!\left(x+z,y+u_{\beta}(z)\right)-f(x,y)-z\frac{\partial f}{\partial
x}(x,y)-u_{\beta}(z)\frac{\partial f}{\partial y}(x,y)\right]c_{\varepsilon}(z)h(z)\,dz,
\end{align}
and where $u_{\beta}(z)$ is as given in \eqref{eq:ubetaz}. The following lemma states the first-order formula which will be used in the sequel. The proof of the lemma is standard and, thus, is deferred to the appendix.
\begin{lemma}\label{lem:Dynkin}
Under Assumptions \ref{assump:UnifBdVol} and \ref{assump:SmthBdLevyDen}, for any $\varepsilon\in(0,1)$ and $f\in C_{b}^{2}(\mathbb{R}^{2})$, we have
\begin{align}\label{eq:Dynkin}
\mathbb{E}\left(f\!\left(X_{t}^{\varepsilon}(x),Y_{t}^{\varepsilon}(x,y)\right)\right)=f(x,y)+t\int_{0}^{1}\mathbb{E}\left(L_{\varepsilon}f\!\left(X_{\alpha
t}^{\varepsilon}(x),Y_{\alpha t}^{\varepsilon}(x,y)\right)\right)d\alpha.
\end{align}
Moreover, there exists a constant $C_{1}>0$, depending on $\beta$, $\varepsilon$, $\|\partial^{i}f\|_{\infty}$, $i=0,1,2$, and $\|\sigma\|_{\infty}$, such that $\|L_{\varepsilon}f\|_{\infty}\leq C_{1}$.
\end{lemma}

\section{Options on the LETF}\label{sec:OptLETFs}

Consider an out-of-the-money (OTM) European call option on the LETF $L$ (with leverage ratio $\beta\in(-\infty,-1]\cup[1,\infty)$), with maturity $t>0$ and strike price $K>e^{x}$. Let $\Pi(t;x,K,\beta)$ denote the time-zero price of such an OTM call option. That is,
\begin{align}\label{eq:OTMOptPriceLETF0}
\Pi(t;x,K,\beta):=\mathbb{E}\left(\left(L_{t}-K\right)^{+}\right)=\mathbb{E}\left(\left({\bf
1}_{\{\tau>t\}}e^{Y_{t}}-K\right)^{+}\right)=\mathbb{E}\left({\bf 1}_{\{\tau>t\}}\left(e^{Y_{t}}-K\right)^{+}\right).
\end{align}
We are interested in the small-maturity behavior of $\Pi(t;x,K,\beta)$ as $t\rightarrow 0$. In light of \eqref{eq:DefaultTimeLETF}, \eqref{eq:DynETF}, and \eqref{eq:DynLETF}, by conditioning on $N([0,t],A^{c})$, we have
\begin{align}\label{eq:OTMOptPriceLETF}
\Pi(t;x,K,\beta)=e^{-t\nu(A^{c})}\mathbb{E}\left(\left(e^{\widetilde{Y}_{t}(x)}-K\right)^{+}\right)=e^{-t\nu(A^{c})}\mathbb{E}\left(\left(e^{Y^{o}_{t}(x,x)}-K\right)^{+}\right),
\end{align}
where $\widetilde{Y}(x)$ and $Y^{o}(x,x)$ are defined in \eqref{eq:DynTildeY} and \eqref{eq:SDEYo}, respectively. Similar to the approach in earlier works (cf.~\cite{FigueroaLopezLuoOuyang:2014}), in order to analyze the small-time asymptotic behavior of the moment of $Y^{o}_{t}(x,x)$, given as on the right-hand side of \eqref{eq:OTMOptPriceLETF}, we take advantage of the decomposition \eqref{eq:DecompZA}, by conditioning on the number of ``big" jumps occurring up to time $t$. More precisely, recalling that $N^{\varepsilon}(A):=(N_{t}^{\varepsilon}(A))_{t\geq 0}$ and $\lambda_{\varepsilon}(A)$ represent, respectively, the jump counting process and the jump intensity of the big-jump component $Z^{\varepsilon,2}(A)$, we have
\begin{align}\label{eq:DecompPi}
\Pi(t;x,K,\beta)=e^{-t\nu(A^{c})}\,e^{-t\lambda_{\varepsilon}(A)}\left(I_{1}+I_{2}+I_{3}\right),
\end{align}
where
\begin{align}\label{eq:I1}
I_{1}(t)=I_{1}(t;x,K,\varepsilon,\beta)&:=\mathbb{E}\left(\left.\left(e^{Y^{o}_{t}(x,x)}-K\right)^{+}\,\right|N_{t}^{\varepsilon}(A)=0\right)=\mathbb{E}\left(\left(e^{Y^{\varepsilon}_{t}(x,x)}-K\right)^{+}\right),\\
\label{eq:I2}
I_{2}(t)=I_{2}(t;x,K,\varepsilon,\beta)&:=t\lambda_{\varepsilon}(A)\,\mathbb{E}\left(\left.\left(e^{Y^{o}_{t}(x,x)}-K\right)^{+}\,\right|N_{t}^{\varepsilon}(A)=1\right),\\
\label{eq:I3}
I_{3}(t)=I_{3}(t;x,K,\varepsilon,\beta)&:=t^{2}\lambda_{\varepsilon}^{2}(A)\sum_{n=2}^{\infty}\frac{\left(\lambda_{\varepsilon}(A)\,t\right)^{n-2}}{n!}\,\mathbb{E}\left(\left.\left(e^{Y^{o}_{t}(x,x)}-K\right)^{+}\,\right|N_{t}^{\varepsilon}(A)=n\right).
\end{align}

\begin{remark}
As in the previous works (cf.~\cite{FigueroaLopezGongHoudre:2012}, \cite{FigueroaLopezLuoOuyang:2014} and~\cite{Leandre:1987}), for $t>0$ small enough, the component $I_{1}(t)$ with no ``big'' jumps is expected to be negligible compared to any power of $t$, while those terms in $I_{2}(t)$ and $I_{3}(t)$, where at least one ``big'' jump is present, are expected to contribute to a polynomial asymptotic expansion of $\Pi$ in powers of $t$.  However, unlike the previous works (cf.~\cite{FigueroaLopezGongHoudre:2012}, \cite{FigueroaLopezLuoOuyang:2014} and~\cite{Leandre:1987}), as we show below, important differences arise when analyzing the asymptotics of the above terms due to (i) the perfectly correlated noise and jump structure, (ii) the boundedness restriction on the jump sizes (Remark \ref{rem:JumpSupport}), as well as (iii) the singularity of the jump coefficients in the model \eqref{eq:SDEYo}-\eqref{eq:SDEXo}.  These differences prevent us from applying the approach in previous works (where either the jumps and the noise are independent, or the dynamic is one-dimensional with unbounded jump size and smooth jump coefficient) to our present setting.
\end{remark}

We begin with the following lemma on the short-maturity asymptotic behavior of $I_{1}(t)$, which shows that, by choosing $\varepsilon>0$ small enough, we can make $I_{1}(t)$ of an arbitrarily large polynomial order in $t$. The proof is similar to those of~\cite[Proposition 3.1]{FigueroaLopezLuoOuyang:2014} and~\cite[Proposition I.4]{Leandre:1987}, with some minor technical differences, and is thus deferred to the appendix.

\begin{lemma}\label{lem:SmallTimeEstI1}
Let Assumptions \ref{assump:UnifBdVol} and \ref{assump:SmthBdLevyDen} be valid, let $K>e^{x}$, and let $\beta\in(-\infty,-1]\cup[1,\infty)$. Then for any $n\in\mathbb{N}$, and any $\varepsilon\in(0,\ln((e^{(\ln K-x)/(2n)}-1)/|\beta|+1)\wedge |\ln(1-\beta^{-1})|\wedge 1)$, there exists $C_{2}>0$, depending on $K$, $x$, $\varepsilon$, $|\beta|$ and $\|\sigma\|_{\infty}$, such that $|I_{1}(t)|\leq C_{2}t^{n}$, for all $t\in[0,e^{-3\beta_{\varepsilon}}]$, where $\beta_{\varepsilon}:=\ln(|\beta|(e^{\varepsilon}-1)+1)$.
\end{lemma}

Next, we will analyze the small-maturity behavior of $I_{2}(t)$, given as in \eqref{eq:I2}. By conditioning on the time of the jump of $Z^{\varepsilon,2}(A)$ and using the Markov property of the pair $(X^{\varepsilon},Y^{\varepsilon})$, $I_{2}(t)$ can be further expressed as
\begin{align}\label{eq:I2Alt}
I_{2}(t)=I_{2}(t;x,K,\varepsilon,\beta)&\,\,=K\lambda_{\varepsilon}(A)\int_{0}^{t}\mathbb{E}\left(G_{t-s}\left(X_{s}^{\varepsilon}(x),Y_{s}^{\varepsilon}(x,x-\ln
K);\varepsilon,\beta\right)\right)ds,
\end{align}
where
\begin{align}\label{eq:FuntG}
G_{t}(\bar{x},\bar{y};\varepsilon,\beta):=\mathbb{E}\left(\left(e^{Y^{\varepsilon}_{t}(\bar{x}+J_{\varepsilon},\bar{y}+u_{\beta}(J_{\varepsilon}))}-1\right)^{+}\right),\quad
t\geq 0,\quad\bar{x},\bar{y}\in\mathbb{R}.
\end{align}
Above, we recall that $u_{\beta}$ is as given in \eqref{eq:ubetaz}, and $J_{\varepsilon}$ is a random variable, independent of $X^{\varepsilon}$ and $Y^{\varepsilon}$, with density \eqref{eq:DenJumpZ2epsA}. Note that
\begin{align}
G_{0}(\bar{x},\bar{y};\varepsilon,\beta)=\mathbb{E}\left(\left(e^{\bar{y}+u_{\beta}(J_{\varepsilon})}-1\right)^{+}\right)
\end{align}
only depends on $\bar{y}$.

\begin{remark}\label{rmk:matt}
The first key step in analyzing the small-maturity expansion of the option price $\Pi(t)$ is to derive the small-maturity asymptotics of the function $G_{t}$, defined above.  The approach taken in~\cite{FigueroaLopezLuoOuyang:2014} is to approximate the function $z\mapsto (e^{z}-1)^{+}$ via a sequence of smooth and bounded functions, the expectations of which, when composed with $Y_{t}^{\varepsilon}(\bar{x}+J_{\varepsilon},\bar{y}+u_{\beta}(J_{\varepsilon}))$, are expanded via the (iterated-type) Dynkin's formula, and provide a uniform bound on the remainder using the theoretical machinery of the flow of diffeomorphisms for SDEs.  However, in our present setting, the approach taken in~\cite{FigueroaLopezLuoOuyang:2014} fails for two reasons that we now proceed to explain.

First, note that the expansion of $G_{t}$ will be plugged into \eqref{eq:I2Alt} to obtain the short-maturity expansion of $I_{2}(t)$, by further expanding the expectations of those coefficients in the expansion of $G_{t}$ composed with $(X_{s}^{\varepsilon}(x),Y_{s}^{\varepsilon}(x,x-\ln K);\varepsilon,\beta)$, using the Dynkin's formula.
This will require that the coefficients in the expansion of $G_{t}$ are smooth (in $\bar{x}$ and $\bar{y}$) and bounded. However, when $\beta\leq -1$, the jump measures of both $X^{o}$ and $Y^{o}$ are supported on half-lines, and the density function $g_{J}$ (see \eqref{eq:DenJumpZ2epsA}) of the size of a ``big'' jump is not smooth at the end points of those half-lines. As a consequence, the coefficients of $G_{t}$ resulting from the Dynkin's formula will be smooth only up to $C^{2}$ (with extra assumption \eqref{eq:AdContLevyDen}), and they will blow up at the end points of half-lines.

Second,the uniform bound on the remainder term in~\cite{FigueroaLopezLuoOuyang:2014} relies heavily on the observation that under the one-dimensional dynamic $X(x):=(X_{t}(x))_{t\geq 0}$ in~\cite{FigueroaLopezLuoOuyang:2014}, the map $x\mapsto X_{t}^{\varepsilon}(x)$ is a diffeomorphism, where $X_{t}^{\varepsilon}(x)$ is the ``small-jump'' part of $X_{t}(x)$, obtained via a similar truncation as above. This important observation is proved using the regularity assumption therein and the fact that the jump measure of $X(x)$ is fully supported and smooth on $\mathbb{R}$. However, in our model, for both $\beta>1$ and $\beta\leq -1$, the mapping $(x,y)\mapsto (X_{t}^{\varepsilon}(x,y),Y_{t}^{\varepsilon}(x,y))$ fails to be a homeomorphism, even in the case of finite jump activity (so that $(X_{t}^{\varepsilon},Y_{t}^{\varepsilon})$ is purely a continuous diffusion) and, thus, the remainder terms would not admit uniform bounds.
\end{remark}

To overcome the two difficulties outlined in Remark \ref{rmk:matt}, we provide a direct approximation of $G_{t}(\bar{x},\bar{y};\varepsilon,\beta)$ by $G_{0}(\bar{x},\bar{y};\varepsilon,\beta)$, up to an error term of order $O(\sqrt{t})$, which is given in the following lemma.
\begin{lemma}\label{lem:SmallTimeExpG}
Let Assumptions \ref{assump:UnifBdVol} and \ref{assump:SmthBdLevyDen} be valid. Let $\beta\in(-\infty,-1]\cup[1,\infty)$ and let
$\varepsilon\in(0,|\ln(1-1/\beta)|\wedge 1)$. Then, there exists a constant $C_{3}>0$, depending only on $\beta$, $\varepsilon$, and
$\|\sigma\|_{\infty}$, such that for any $\bar{x},\bar{y}\in\mathbb{R}$, and any $t\geq 0$,
\begin{align}
\left|G_{t}(\bar{x},\bar{y};\varepsilon,\beta)-{G_{0}(\bar{x},\bar{y};\varepsilon,\beta)}\right|\leq C_{3}\,\mathbb{E}\left(e^{u_{\beta}(J_{\varepsilon})}\right)e^{\bar{y}}\sqrt{t}.
\end{align}
\end{lemma}

\noindent
\textbf{Proof:} For ease of notation, we simply write $G_{0}(\bar{y};\varepsilon,\beta)$ in stead of $G_{0}(\bar{x},\bar{y};\varepsilon,\beta)$ throughout the proof. Also, since we fix any $\beta\in(-\infty,-1]\cup[1,\infty)$ and $\varepsilon\in(0,|\ln(1-1/\beta)|\wedge 1)$ throughout the proof, we will omit the parameters $\varepsilon$ and $\beta$ in $G_{t}(\bar{x},\bar{y};\varepsilon,\beta)$ and $G_{0}(\bar{x},\bar{y};\varepsilon,\beta)$. We will first look for a bound on
\begin{align}\label{eq:Etxy} \mathcal{E}_{t}(\tilde{x},\tilde{y})=\mathcal{E}_{t}(\tilde{x},\tilde{y};\varepsilon,\beta):=\mathbb{E}\left(\left(e^{Y^{\varepsilon}_{t}(\tilde{x},\tilde{y})}-1\right)^{+}\right)-\left(e^{\tilde{y}}-1\right)^{+},\quad\tilde{x},\tilde{y}\in\mathbb{R},
\end{align}
since
\begin{align}\label{eq:RelEtGt}
\left|G_{t}(\bar{x},\bar{y})-G_{0}(\bar{y})\right|\leq\mathbb{E}\left|\mathcal{E}_{t}\left(\bar{x}+J_{\varepsilon},\bar{y}+u_{\beta}(J_{\varepsilon}\right))\right|.
\end{align}
To begin with, we decompose $Y^{\varepsilon}(\tilde{x},\tilde{y})$ as
\begin{align}\label{DcmpYepsilon}
Y_{t}^{\varepsilon}(\tilde{x},\tilde{y})=\tilde{y}+d_{\varepsilon}t+R_{t}^{\varepsilon}:=\tilde{y}+d_{\varepsilon}t+R_{t}^{\varepsilon,1}+R_{t}^{\varepsilon,2},\quad
t\geq 0,
\end{align}
where
\begin{align}
R_{t}^{\varepsilon,1}&:=\beta\int_{0}^{t}\sigma\!\left(X_{s}^{\varepsilon}(\tilde{x})\right)d\widetilde{W}_{s}-\frac{\beta^{2}}{2}\int_{0}^{t}\sigma^{2}\!\left(X_{s}^{\varepsilon}(\tilde{x})\right)ds\\
R_{t}^{\varepsilon,2}&:=\int_{0}^{t}\int_{A_{0}}u_{\beta}(z)\,\widetilde{M}^{\varepsilon,1}_{A}(ds,dz)-t\int_{A_{0}}\left[\beta\left(e^{z}-1\right)-u_{\beta}(z)\right]c_{\varepsilon}(z)h(z)\,dz,\\
\label{Dfndepsilon}
d_{\varepsilon}&:=\nu(A^{c})-\int_{A}\beta\left(e^{z}-1\right)\left(1-c_{\varepsilon}(z)\right)h(z)\,dz.
\end{align}
Let $\widetilde{R}^{\varepsilon}_{t}:=e^{R_{t}^{\varepsilon}}$, $t\geq 0$. By It\^{o}'s formula,
\begin{align}
d\widetilde{R}^{\varepsilon}_{t}=\beta\widetilde{R}^{\varepsilon}_{t}\sigma\!\left(X_{t}^{\varepsilon}(\tilde{x})\right)d\widetilde{W}_{t}+\widetilde{R}^{\varepsilon}_{t-}\int_{A_{0}}\beta\left(e^{z}-1\right)\widetilde{M}^{\varepsilon,1}_{A}(dt,dz).
\end{align}
Hence, $\widetilde{R}^{\varepsilon}:=(\widetilde{R}^{\varepsilon}_{t})_{t\geq 0}$ is an $\Fb$-local martingale under $\mathbb{P}$. To find the bound
for \eqref{eq:Etxy}, we need that $\widetilde{R}^{\varepsilon}$ is an $(\mathscr{F}_{t})_{t\geq 0}$-martingale under $\mathbb{P}$. In light of Assumption \ref{assump:UnifBdVol}, we need to show that, for any $t\geq 0$,
\begin{align}\label{eq:MartCondREps}
\int_{0}^{t}\mathbb{E}\left(e^{2R_{s}^{\varepsilon}}\right)ds<\infty,\quad\int_{0}^{t}\int_{A_{0}}\mathbb{E}\left(e^{2R_{s}^{\varepsilon}}\right)\left(e^{z}-1\right)^{2}c_{\varepsilon}(z)h(z)\,dz\,ds<\infty.
\end{align}
Since $\int_{A_{0}}\left(e^{z}-1\right)^{2}c_{\varepsilon}(z)h(z)\,dz<\infty$, it suffices to bound the integrand in the first integral in \eqref{eq:MartCondREps}. Now, for any $s\geq 0$, by the Cauchy-Schwarz inequality,
\begin{align}\label{eq:DecompExpREps}
\mathbb{E}\left(e^{2R_{s}^{\varepsilon}}\right)\leq\left(\mathbb{E}\left(e^{4R_{s}^{\varepsilon,1}}\right)\right)^{1/2}\left(\mathbb{E}\left(e^{4R_{s}^{\varepsilon,2}}\right)\right)^{1/2}.
\end{align}
For the expectation with respect to $R_{s}^{\varepsilon,1}$, by Assumption \ref{assump:UnifBdVol},
\begin{align}
\mathbb{E}\left(e^{4R_{s}^{\varepsilon,1}}\right)&=\mathbb{E}\left(\exp\left(4\beta\int_{0}^{s}\sigma\left(X_{u}^{\varepsilon}(\tilde{x})\right)d\widetilde{W}_{u}-2\beta^{2}\int_{0}^{s}\sigma^{2}\left(X_{u}^{\varepsilon}(\tilde{x})\right)du\right)\right)\nonumber\\
&=\mathbb{E}\left(\exp\left(6\beta^{2}\int_{0}^{s}\sigma^{2}\left(X_{u}^{\varepsilon}(\tilde{x})\right)du\right)\cdot\exp\left(4\beta\int_{0}^{s}\sigma\left(X_{u}^{\varepsilon}(\tilde{x})\right)d\widetilde{W}_{u}-8\beta^{2}\int_{0}^{s}\sigma^{2}\left(X_{u}^{\varepsilon}(\tilde{x})\right)du\right)\right)\nonumber\\
\label{eq:EstExpR1Eps} &\leq
e^{6\beta^{2}\|\sigma\|_{\infty}^{2}s}\cdot\mathbb{E}\left(\exp\left(4\beta\int_{0}^{s}\sigma\!\left(X_{u}^{\varepsilon}(\tilde{x})\right)d\widetilde{W}_{u}-8\beta^{2}\int_{0}^{s}\sigma^{2}\!\left(X_{u}^{\varepsilon}(\tilde{x})\right)du\right)\right)=e^{6\beta^{2}\|\sigma\|_{\infty}^{2}s}.
\end{align}
For the expectation with respect to $R_{s}^{\varepsilon,2}$, we have
\begin{align}
\mathbb{E}\left(e^{4R_{s}^{\varepsilon,2}}\right)&=\mathbb{E}\left(\exp\left(\int_{0}^{s}\!\int_{A_{0}}4u_{\beta}(z)\,\widetilde{M}^{\varepsilon,1}_{A}(du,dz)-s\int_{A_{0}}\left\{\left[\beta\left(e^{z}-1\right)+1\right]^{4}-1-4u_{\beta}(z)\right\}c_{\varepsilon}(z)h(z)\,dz\right)\right)\\
&\quad\quad\,\,\times\exp\left(s\int_{A_{0}}\left\{\left[\beta\left(e^{z}-1\right)+1\right]^{4}-1-\beta\left(e^{z}-1\right)-3u_{\beta}(z)\right\}c_{\varepsilon}(z)h(z)\,dz\right)\\
\label{eq:EstExpR2Eps}
&=\exp\left(s\int_{A_{0}}\left\{\left[\beta\left(e^{z}-1\right)+1\right]^{4}-1-4\beta\left(e^{z}-1\right)\right\}c_{\varepsilon}(z)h(z)\,dz\right).
\end{align}
Above, we note that
\begin{align}
\left[\beta\left(e^{z}-1\right)+1\right]^{4}-1-4\beta\left(e^{z}-1\right)\sim\beta^{4}z^{4}+4\beta^{3}z^{3}+6\beta^{2}z^{2},\quad
z\rightarrow 0,
\end{align}
so that it is integrable in a neighborhood of the origin with respect to $\nu$. Combining \eqref{eq:DecompExpREps}, \eqref{eq:EstExpR1Eps}, and
\eqref{eq:EstExpR2Eps}, we have shown \eqref{eq:MartCondREps}. Therefore, $\widetilde{R}^{\varepsilon}$ is an $\Fb$-martingale under $\mathbb{P}$ and, thus,
\begin{align}
\mathbb{E}\left(\widetilde{R}_{t}^{\varepsilon}\right)=\mathbb{E}\left(e^{R_{t}^{\varepsilon}}\right)=1,\quad\text{for any }\,t\geq 0.
\end{align}
Coming back to the estimation of \eqref{eq:Etxy}, we have
\begin{align}
\left|\mathcal{E}_{t}(\tilde{x},\tilde{y})\right|\leq
e^{\tilde{y}}\,\mathbb{E}\left(e^{R_{t}^{\varepsilon}}\left|e^{d_{\varepsilon}t}-1\right|\right)+e^{\tilde{y}}\,\mathbb{E}\left(\left|e^{R_{t}^{\varepsilon}}-1\right|\right).
\end{align}
For the second term above, using \eqref{eq:EstExpR1Eps} and \eqref{eq:EstExpR2Eps}, we have
\begin{align}
\mathbb{E}\left(\left|e^{R_{t}^{\varepsilon}}-1\right|^{2}\right)&=\mathbb{E}\left(e^{2R_{t}^{\varepsilon}}\right)-1\leq\left(\mathbb{E}\left(e^{4R_{t}^{\varepsilon,1}}\right)\right)^{1/2}\left(\mathbb{E}\left(e^{4R_{t}^{\varepsilon,2}}\right)\right)^{1/2}-1\\
&\leq\exp\left(t\left(3\beta^{2}\|\sigma\|_{\infty}^{2}+\frac{1}{2}\!\int_{A_{0}}\!\left\{\left[\beta\left(e^{z}\!-1\right)\!+\!1\right]^{4}\!-\!1\!-\!\beta\left(e^{z}\!-1\right)\!-\!3u_{\beta}(z)\right\}c_{\varepsilon}(z)h(z)\,dz\right)\right)-1.
\end{align}
Letting
\begin{align}\label{eq:Constc}
c=c(\varepsilon,\beta,\|\sigma\|_{\infty}):=3\beta^{2}\|\sigma\|_{\infty}^{2}+\frac{1}{2}\int_{A_{0}}\left\{\left[\beta\left(e^{z}-1\right)+1\right]^{4}-1-\beta\left(e^{z}-1\right)-3u_{\beta}(z)\right\}c_{\varepsilon}(z)h(z)\,dz,
\end{align}
we obtain that
\begin{align}\label{Eq:BndC2}
\left|\mathcal{E}_{t}(\tilde{x},\tilde{y})\right|\leq
e^{\tilde{y}}\,\mathbb{E}\left(e^{R_{t}^{\varepsilon}}\left|e^{d_{\varepsilon}t}-1\right|\right)+e^{\tilde{y}}\,\sqrt{e^{ct}-1}\leq
C_{3}\,e^{\tilde{y}}\sqrt{t},
\end{align}
where $C_{3}>0$ is a constant depending only on $\beta$, $\varepsilon$ and $\|\sigma\|_{\infty}$. The lemma follows immediately from \eqref{eq:RelEtGt} and the above bound on $\mathcal{E}_{t}(\tilde{x},\tilde{y};\varepsilon,\beta)$.\hfill $\Box$

\medskip
The second key step for analyzing the small-time asymptotic behavior of $I_{2}$ is to apply the Dynkin's formula to
$\mathbb{E}\left(G_{0}\left(X_{s}^{\varepsilon}(x),Y_{s}^{\varepsilon}(x,x-\ln K);\varepsilon,\beta\right)\right)$. Note that
\begin{align}\label{eq:G0TailProb}
G_{0}(\bar{x},\bar{y};\varepsilon,\beta)=\mathbb{E}\left(\left(e^{\bar{y}+u_{\beta}(J_{\varepsilon})}-1\right)^{+}\right)=\int_{1}^{\infty}\mathbb{P}\left(u_{\beta}(J_{\varepsilon})>\ln z-\bar{y}\right)dz,
\end{align}
where, again, $u_{\beta}$ is as defined in \eqref{eq:ubetaz}. Therefore, in order to apply Lemma \ref{lem:Dynkin} to the integrand above (which is clearly bounded), we need to establish the smoothness of
\begin{align}\label{eq:FuntH0}
(\bar{x},\bar{y})\mapsto H_{0}(\bar{x},\bar{y};z,\varepsilon,\beta):=\mathbb{P}\left(u_{\beta}(J_{\varepsilon})>\ln z-\bar{y}\right),
\end{align}
for each fixed $z\geq 1$, which is shown in the following lemma. The reason why we write $H_{0}(\bar{x},\bar{y};z,\varepsilon,\beta)$ as a function of $\bar{x}$, even though it only depends on $\bar{y}$, is because we eventually need to apply the Dynkin's formula to $\mathbb{E}\left(H_{0}\left(X_{s}^{\varepsilon}(x),Y_{s}^{\varepsilon}(x,x-\ln
K);\varepsilon,\beta\right)\right)$. However, in what follows, we shall often omit $\bar{x}$ when writing the function $H_{0}$. Clearly, we only need to check the smoothness of $H_{0}$ with respect to $\bar{y}$.
\begin{lemma}\label{lem:SmoothH0}
Let Assumption \ref{assump:SmthBdLevyDen} be valid. Let $\beta\in(-\infty,-1]\cup[1,\infty)$ and let $\varepsilon\in(0,|\ln(1-1/\beta)|\wedge 1)$.
For any fixed $z\geq 1$, let $H_{0}(\,\cdot\,;z,\varepsilon,\beta)$ be defined as in \eqref{eq:FuntH0}. Then, for $\beta\geq 1$, $H_{0}(\,\cdot\,;z,\varepsilon,\beta)\in C_{b}^{2}(\mathbb{R})$ always, while, for $\beta\leq -1$, $H_{0}(\,\cdot\,;z,\varepsilon,\beta)\in
C_{b}^{2}(\mathbb{R})$, provided that the L\'{e}vy density $h$ satisfies the following additional condition:
\begin{align}\label{eq:AdContLevyDen}
\lim_{y\rightarrow -\infty}e^{-ky}h^{(k)}(y)=0,\quad\text{for }\,k=1,2.
\end{align}
\end{lemma}

\noindent
\textbf{Proof:} We first assume that $\beta=1$, then $A=\mathbb{R}$ and $u_{\beta}(x)=x$, and hence $H_{0}(\,\cdot\,;z,\varepsilon,1)\in
C_{b}^{2}(\mathbb{R})$ since the density of $J_{\varepsilon}$ satisfies $g_{J}(\,\cdot\,;\varepsilon,\beta)\in C_{b}^{2}(\mathbb{R})$ by
Assumption \ref{assump:SmthBdLevyDen}-(iii). Next, assume that $\beta>1$. Denote the density of $U_{\varepsilon}:=u_{\beta}(J_{\varepsilon})$ by
$g_{U}(\,\cdot\,;\varepsilon,\beta)$. Using Assumption \ref{assump:SmthBdLevyDen}-(iii) once again, we find that
\begin{align}
H_{0}'(\bar{y};z,\varepsilon,\beta)=g_{U}(\ln
z-\bar{y};\varepsilon,\beta)=g_{J}'\!\left(\ln\left(\frac{ze^{-\bar{y}}-1}{\beta}+1\right);\varepsilon,\beta\right)\frac{ze^{-\bar{y}}}{ze^{-\bar{y}}-1+\beta}\in C_{b}^{1}(\mathbb{R}).
\end{align}
Hence, we still have $H_{0}(\,\cdot\,;z,\varepsilon,\beta)\in C_{b}^{2}(\mathbb{R})$.

Finally, we study the regularity of $\bar{y}\mapsto H_{0}(\bar{y};z,\varepsilon,\beta)$ when $\beta\leq -1$. Clearly,
$H_{0}(\,\cdot\,;z,\varepsilon,\beta)\in C_{b}(\mathbb{R})$, and by Remark \ref{rem:JumpSupport} (the domains $A$ and $u_{\beta}(A)$ for $\beta\leq
-1$) and Assumption \ref{assump:SmthBdLevyDen} (so that $J_{\varepsilon}$ has no atom), for $\bar{y}\leq\ln z-\ln(1-\beta)$, $H_{0}(\bar{y};z,\varepsilon,\beta)=0$. Now, for $\bar{y}>{\ln z}-\ln(1-\beta)$, by \eqref{eq:DenJumpZ2epsA}, we have
\begin{align}
H_{0}'(\bar{y};z,\varepsilon,\beta)\!&=-\frac{1}{\lambda_{\varepsilon}(A)}\,c_{\varepsilon}'\!\left(\ln\left(\frac{ze^{-\bar{y}}-1}{\beta}+1\right)\right)h\!\left(\ln\left(\frac{ze^{-\bar{y}}-1}{\beta}+1\right)\right)\frac{ze^{-\bar{y}}}{ze^{-\bar{y}}-1+\beta}\\
&\quad\,+\frac{1}{\lambda_{\varepsilon}(A)}\left(1\!-\!c_{\varepsilon}\!\left(\!\ln\!\left(\frac{ze^{-\bar{y}}\!-\!1}{\beta}\!+\!1\right)\!\right)\!\right)h'\!\left(\!\ln\!\left(\frac{ze^{-\bar{y}}\!-\!1}{\beta}\!+\!1\right)\!\right)\frac{ze^{-\bar{y}}}{ze^{-\bar{y}}\!-\!1\!+\!\beta}\\
\label{eq:DenUepsbeta} &=:g_{U,1}(\bar{y};z,\varepsilon,\beta)+g_{U,2}(\bar{y};z,\varepsilon,\beta).
\end{align}
Observe that $c_{\varepsilon}'$ is supported on $[-\varepsilon,-\varepsilon/2]\cup[\varepsilon/2,\varepsilon]$, and thus
$g_{U,1}(\,\cdot\,;z,\varepsilon,\beta)$ is supported on $[\ln z-u_{\beta}(\varepsilon/2),\ln z-u_{\beta}(\varepsilon)]\cup[\ln
z-u_{\beta}(-\varepsilon/2),\ln z-u_{\beta}(-\varepsilon)]$, which clearly excludes a neighborhood of the singular point $\bar{y}=\ln z$ of
$h(\ln((ze^{-\bar{y}}-1)/\beta+1))$, as well as the singular point $\bar{y}=\ln z-\ln(1-\beta)$ of $ze^{-\bar{y}}/(ze^{-\bar{y}}-1+\beta)$. Hence,
$g_{U,1}(\,\cdot\,;z,\varepsilon,\beta)\in C_{b}^{2}(\ln z-\ln(1-\beta),\infty)$ by Assumption \ref{assump:SmthBdLevyDen}-(iii), and
moreover,
\begin{align}
\lim_{\bar{y}\downarrow\ln z-\ln(1-\beta)}g_{U,1}^{(n)}(\bar{y};z,\varepsilon,\beta)=0,\quad\text{for all }\,n=1,2.
\end{align}
For $g_{U,2}$, we first note, by \eqref{eq:AdContLevyDen} with $k=1$, that
\begin{align}
\lim_{\bar{y}\downarrow\ln
z-\ln(1-\beta)}h'\!\left(\ln\left(\frac{ze^{-\bar{y}}-1}{\beta}+1\right)\right)\frac{ze^{-\bar{y}}}{ze^{-\bar{y}}-1+\beta}=\lim_{y\rightarrow-\infty}\beta
e^{-y}h'(y)\left[\beta\left(e^{y}-1\right)+1\right]=0,
\end{align}
and thus,
\begin{align}
\lim_{\bar{y}\downarrow\ln z-\ln(1-\beta)}g_{U,2}(\bar{y};z,\varepsilon,\beta)=0.
\end{align}
Moreover, for $\bar{y}>\ln z-\ln(1-\beta)$,
\begin{align}
g_{U,2}'(\bar{y};z,\varepsilon,\beta)&=-\frac{1}{\lambda_{\varepsilon}(A)}\,c_{\varepsilon}'\!\left(\ln\left(\frac{ze^{-\bar{y}}-1}{\beta}+1\right)\right)h'\!\left(\ln\left(\frac{ze^{-\bar{y}}-1}{\beta}+1\right)\right)\frac{z^{2}e^{-2\bar{y}}}{\left(ze^{-\bar{y}}-1+\beta\right)^{2}}\\
&\quad\,+\frac{1}{\lambda_{\varepsilon}(A)}\left(1-c_{\varepsilon}\!\left(\ln\left(\frac{ze^{-\bar{y}}-1}{\beta}+1\right)\right)\right)h''\!\left(\ln\left(\frac{ze^{-\bar{y}}-1}{\beta}+1\right)\right)\frac{z^{2}e^{-2\bar{y}}}{\left(ze^{-\bar{y}}-1+\beta\right)^{2}}\\
&\quad\,+\frac{1}{\lambda_{\varepsilon}(A)}\left(1-c_{\varepsilon}\!\left(\ln\left(\frac{ze^{-\bar{y}}-1}{\beta}+1\right)\right)\right)h'\!\left(\ln\left(\frac{ze^{-\bar{y}}-1}{\beta}+1\right)\right)\frac{ze^{-\bar{y}}}{ze^{-\bar{y}}-1+\beta}.
\end{align}
An argument similar to the analysis of $g_{U,1}$ shows that the first term above belongs to $C_{b}^{1}(\mathbb{R})$, and that its right limit at
$\ln z-\ln(1-\beta)$ equals 0. Also, as $1-c_{\varepsilon}$ is supported on $(-\infty,-\varepsilon/2)\cup(\varepsilon/2,\infty)$, the second term in
the decomposition of $g_{U,2}$ above is supported on $(\ln z-\ln(1-\beta),\ln z-u_{\beta}(-\varepsilon/2))\cup(\ln
z-u_{\beta}(\varepsilon/2),\infty)$, which excludes a neighborhood of the singular point $\bar{y}=\ln z$ of $h'(\ln((ze^{-\bar{y}}-1)/\beta+1))$.
Moreover, \eqref{eq:AdContLevyDen} with $k=2$ ensures that the second term in the decomposition of $g_{U,2}$ above has a right limit $0$ at $\ln
z-\ln(1-\beta)$. Therefore, this second term belongs to $C_{b}(\ln z-\ln(1-\beta),\infty)$ with a right limit equal to $0$ at $\ln z-\ln(1-\beta)$.
Similarly, we can show that the third term in the decomposition of $g_{U,2}$ above belongs to $C_{b}(\ln z-\ln(1-\beta),\infty)$ with a right limit
equal to $0$ at $\ln z-\ln(1-\beta)$. To sum up, we have shown that $H_{0}(\,\cdot\,;z,\varepsilon,\beta)\in C_{b}^{2}(\ln z-\ln(1-\beta),\infty)$
such that $H^{(k)}(\,\cdot\,;z,\varepsilon,\beta)$ has a right limit $0$ at $\bar{y}=\ln z-\ln(1-\beta)$, for $k=0,1,2$, and hence
$H_{0}(\,\cdot\,;z,\varepsilon,\beta)\in C_{b}^{2}(\mathbb{R})$. The proof is now complete.\hfill $\Box$

\medskip
It remains to analyze the behavior of $I_{3}(t)$, given as in \eqref{eq:I3}, which is the content of the next lemma. The proof is a nontrivial generalization of that of~\cite[Lemma 6.1]{FigueroaLopezLuoOuyang:2014} to our two-dimensional correlated model with some additional technical issues, and is presented in the appendix.
\begin{lemma}\label{lem:SmallTimeEstI3}
Let Assumptions \ref{assump:UnifBdVol} and \ref{assump:SmthBdLevyDen} be valid. With the notation given as in Section
\ref{subsec:Notations}, there exists a constant $C_{4}>0$, depending on $\varepsilon$, $\beta$ and $\|\sigma\|_{\infty}$, such that
\begin{align}
\sup_{n\in\mathbb{N},t\in[0,1]}\frac{1}{n!}\int_{0}^{\infty}\mathbb{P}\left(\left.\left|Y_{t}^{o}(x,x)-x\right|>\ln
z\,\right|N_{t}^{\varepsilon}(A)=n\right)dz\leq C_{4}<\infty.
\end{align}
\end{lemma}

Using the above four lemmas, we are now in a position to state and prove the main result of this section.
\begin{theorem}\label{thm:SmallTimeOTMCallOptLETF}
Let $\Pi(t;x,K,\beta)$ be the time-zero price of a European call option on the LETF $L$, where $L_{0}=e^{x}$, with strike price $K>e^{x}$ and maturity $t$. Suppose Assumption \ref{assump:UnifBdVol} and Assumption \ref{assump:SmthBdLevyDen} are valid. Then, for $\beta\geq 1$,
\begin{align}\label{eq:1stOrderAsyOTMCallOptLETF}
\Pi(t;x,K,\beta)=b_{1}t+o(t),\quad t\rightarrow 0,
\end{align}
where
\begin{align}\label{eq:1stCoefOTMCallOptLETF1}
b_{1}=b_{1}(x,K,\beta):=\int_{\ln((Ke^{-x}-1+\beta)/\beta)}^{\infty}\left[\beta e^{x+z}+(1-\beta)e^{x}-K\right]h(z)\,dz.
\end{align}
When $\beta\in(-\infty,-1]\cap(-\infty,1-Ke^{-x})$, \eqref{eq:1stOrderAsyOTMCallOptLETF} remains valid under the additional condition \eqref{eq:AdContLevyDen}, with
\begin{align}\label{eq:1stCoefOTMCallOptLETF2}
b_{1}=b_{1}(x,K,\beta):=\int_{-\infty}^{\ln((Ke^{-x}-1+\beta)/\beta)}\left[\beta e^{x+z}+(1-\beta)e^{x}-K\right]h(z)\,dz.
\end{align}
\end{theorem}

We will provide a more thorough discussion of the error term $o(t)$ that appears in \eqref{eq:1stOrderAsyOTMCallOptLETF} in Remark \ref{rmk:error}.
\\[0.5em]
\noindent
\textbf{Proof:} We follow the notations introduced in \eqref{eq:DecompPi}-\eqref{eq:FuntG} with a fixed $\varepsilon\in(0,((\ln
K-x)/2)\wedge|\ln(1-1/\beta)|\wedge|\ln((Ke^{-x}-1+\beta)/\beta)|\wedge 1)$. By Lemma \ref{lem:SmallTimeEstI1}, $I_{1}(t)=O(t^{2})$ for all $t\in(0,1]$. Next, by \eqref{eq:ExpToTailProb}, we can rewrite the expression \eqref{eq:I3} of $I_{3}(t)$ as
\begin{align} I_{3}(t)=t^{2}\lambda_{\varepsilon}^{2}(A)e^{x}\sum_{n=2}^{\infty}\frac{\left(\lambda_{\varepsilon}(A)\,t\right)^{n-2}}{n!}\int_{K/e^{x}}^{\infty}\mathbb{P}\left(\left.Y^{o}_{t}(x,x)-x>\ln
z\,\right|N_{t}^{\varepsilon}(A)=n\right)dz.
\end{align}
Hence, by Lemma \ref{lem:SmallTimeEstI3}, $I_{3}(t)=O(t^{2})$ as $t\rightarrow 0$. Therefore, we obtain that
\begin{align}\label{eq:LimitPiLimitI2}
\lim_{t\rightarrow 0}\frac{1}{t}\,\Pi(t;x,K,\beta)=\lim_{t\rightarrow 0}\frac{I_{2}(t)}{t},
\end{align}
whenever the latter limit exists. To study the above limit, we first rewrite the expression \eqref{eq:I2Alt} of $I_{2}(t)$ as
\begin{align}
I_{2}(t)&=K\lambda_{\varepsilon}(A)\int_{0}^{t}\mathbb{E}\left(G_{0}\left(X_{s}^{\varepsilon}(x),Y_{s}^{\varepsilon}(x,x-\ln
K);\varepsilon,\beta\right)\right)ds\\
&\quad\,+K\lambda_{\varepsilon}(A)\int_{0}^{t}\left(\mathbb{E}\left(G_{t-s}\left(X_{s}^{\varepsilon}(x),Y_{s}^{\varepsilon}(x,x-\ln
K);\varepsilon,\beta\right)\right)-\mathbb{E}\left(G_{0}\left(X_{s}^{\varepsilon}(x),Y_{s}^{\varepsilon}(x,x-\ln
K);\varepsilon,\beta\right)\right)\right)ds.
\end{align}
By Lemma \ref{lem:SmallTimeExpG}, the second term above is such that
\begin{align}
&\int_{0}^{t}\left|\mathbb{E}\left(G_{t-s}\left(X_{s}^{\varepsilon}(x),Y_{s}^{\varepsilon}(x,x-\ln
K);x,K,\varepsilon,\beta\right)\right)-\mathbb{E}\left(G_{0}\left(X_{s}^{\varepsilon}(x),Y_{s}^{\varepsilon}(x,x-\ln
K);x,K,\varepsilon,\beta\right)\right)\right|ds\\
\label{Eq:DfnC4} &\quad\,\leq C_{3}\,\mathbb{E}\left(e^{u_{\beta}(J)}\right)\int_{0}^{t}\mathbb{E}\left(e^{Y_{s}^{\varepsilon}(x,x-\ln
K)}\right)\sqrt{t-s}\,ds=C_{3}\,\mathbb{E}\left(e^{u_{\beta}(J)}\right)\frac{e^{x}}{K}\int_{0}^{t}e^{\overline{C}_{3}s}\sqrt{t-s}\,ds,
\end{align}
where $\overline{C}_{3}>0$ is a constant depending on $\beta$, $\varepsilon$ and $\|\sigma\|_{\infty}$. Indeed, as shown in the proof of Lemma \ref{lem:SmallTimeExpG}, we can take $\overline{C}_{3}=d_{\varepsilon}$ as defined in \eqref{DcmpYepsilon}-\eqref{Dfndepsilon}. Together with \eqref{eq:G0TailProb} and \eqref{eq:FuntH0}, \eqref{eq:LimitPiLimitI2} becomes
\begin{align}
\lim_{t\rightarrow 0}\frac{1}{t}\,\Pi(t;x,K,\beta)&=K\lambda_{\varepsilon}(A)\cdot\lim_{t\rightarrow
0}\frac{1}{t}\int_{0}^{t}\mathbb{E}\left(G_{0}\left(X_{s}^{\varepsilon}(x),Y_{s}^{\varepsilon}(x,x-\ln K);\varepsilon,\beta\right)\right)ds\\
\label{eq:LimitPiLimitH01} &=\lambda_{\varepsilon}(A)\cdot\lim_{t\rightarrow
0}\frac{1}{t}\int_{K}^{\infty}\int_{0}^{t}{\mathbb{E}\left(H_{0}\left(X_{s}^{\varepsilon}(x),Y_{s}^{\varepsilon}(x,x);z,\varepsilon,\beta\right)\right)}ds\,dz.
\end{align}
Note that, for any $t\in[0,1]$, there exists a constant $C_{5}>0$, depending only on $\beta$, $\varepsilon$ and $\|\sigma\|_{\infty}$, such that
\begin{align}
\frac{1}{t}\int_{0}^{t}\mathbb{E}\left(H_{0}\left(X_{s}^{\varepsilon}(x),Y_{s}^{\varepsilon}(x,x);z,\varepsilon,\beta\right)\right)ds&=\frac{1}{t}\int_{0}^{t}\mathbb{P}\left(Y_{s}^{\varepsilon}(x,x)+u_{\beta}(J_{\varepsilon})>\ln
z\right)ds\\
&\leq z^{-1-\delta}\mathbb{E}\left(e^{(1+\delta)u_{\beta}(J)}\right)\frac{1}{t}\int_{0}^{t}\mathbb{E}\left(e^{(1+\delta)Y_{s}^{\varepsilon}(x,x)}\right)ds\\
&\leq z^{-1-\delta}\mathbb{E}\left(e^{(1+\delta)u_{\beta}(J)}\right)e^{C_{5}},
\end{align}
where $\delta>0$ is as defined in Assumption \eqref{assump:SmthBdLevyDen}-(i), which guarantees that $\mathbb{E}\left(e^{(1+\delta)u_{\beta}(J)}\right)<\infty$, when $\beta\geq 1$. Since clearly the above upper bound is integrable with respect to $z$ over $[K,\infty)$, we can apply the dominated convergence theorem to \eqref{eq:LimitPiLimitH01} to get
\begin{align}\label{eq:LimitPiLimitH02}
\lim_{t\rightarrow 0}\frac{1}{t}\,\Pi(t;x,K,\beta)=\lambda_{\varepsilon}(A)\int_{K}^{\infty}\left(\lim_{t\rightarrow
0}\frac{1}{t}\int_{0}^{t}{\mathbb{E}\left(H_{0}\left(X_{s}^{\varepsilon}(x),Y_{s}^{\varepsilon}(x,x);z,\varepsilon,\beta\right)\right)}ds\right)dz,
\end{align}
whenever the latter limit exists. By Lemma \ref{lem:SmoothH0}, the Dynkin's formula (Lemma \ref{lem:Dynkin}) is applicable to $H_{0}(\,\cdot\,;z,\varepsilon,\beta)$. Hence,
\begin{align}
\mathbb{E}\left(H_{0}\left(X_{s}^{\varepsilon}(x),Y_{s}^{\varepsilon}(x,x);z,\varepsilon,\beta\right)\right)=\mathbb{P}\left(u_{\beta}(J_{\varepsilon})>\ln
z-x\right)+s\int_{0}^{1}{\mathbb{E}\left(L_{\varepsilon}H_{0}\left(X_{\alpha s}^{\varepsilon}(x),Y_{\alpha s}^{\varepsilon}(x,x);z,\varepsilon,\beta\right)\right)}d\alpha,
\end{align}
and, moreover, $L_{\varepsilon}H_{0}$ has a finite bound (depending on $x$, $z$, $\varepsilon$ and $\beta$). Therefore, we deduce from
\eqref{eq:LimitPiLimitH02} and the dominated convergence theorem that
\begin{align}
\lim_{t\rightarrow 0}\frac{1}{t}\,\Pi(t;x,K,\beta)=e^{x}\lambda_{\varepsilon}(A)\int_{K/e^{x}}^{\infty}\mathbb{P}\left(u_{\beta}(J_{\varepsilon})>\ln
z\right)dz=\lambda_{\varepsilon}(A)\,\mathbb{E}\left(\left(e^{u_{\beta}(J_{\varepsilon})+x}-K\right)^{+}\right),
\end{align}
which can be rewritten in terms of \eqref{eq:1stCoefOTMCallOptLETF1} and \eqref{eq:1stCoefOTMCallOptLETF2} for $\beta\geq 1$ and $\beta\leq -1$, respectively.\hfill $\Box$

\medskip
Note, by the put-call parity, we have
\begin{align}
\mathbb{E}\left(\left(K-L_{t}\right)^{+}\right)=\mathbb{E}\left(\left(L_{t}-K\right)^{+}\right)+\mathbb{E}\left(K-L_{t}\right)=\mathbb{E}\left(\left(L_{t}-K\right)^{+}\right)+K-e^{x}.
\end{align}
Therefore, we obtain the following result for the corresponding in-the-money (ITM) European put option on the LETF.
\begin{corollary}\label{cor:SmallTimeITMPutOptLETF}
Let $\Theta(t;x,K,\beta)$ be the time-zero price of a European put option on the LETF $L$ (with leverage ratio $\beta$), where $L_{0}=e^{x}$, with strike price $K$ and maturity $t$. Under conditions of Theorem \ref{thm:SmallTimeOTMCallOptLETF},
\begin{align}
\Theta(t;x,K,\beta)=K-e^{x}+b_{1}t+o(t),\quad t\rightarrow 0,
\end{align}
where $b_{1}=b_{1}(x,K,\beta)$ is given by \eqref{eq:1stCoefOTMCallOptLETF1} and \eqref{eq:1stCoefOTMCallOptLETF2}, respectively, for the
cases $\beta\geq 1$ and $\beta\in(-\infty,-1]\cap(-\infty,1-Ke^{-x})$.
\end{corollary}

\begin{remark}
It is worth pointing out the following consequences and remarks:
\begin{itemize}
\item Recalling that $L_{0}=S_{0}=e^{x}$, the coefficients \eqref{eq:1stCoefOTMCallOptLETF1} and \eqref{eq:1stCoefOTMCallOptLETF2} can be written in a more appealing form:
	\begin{align}\label{eq:1stCoefOTMCallOptLETFGen}
    b_{1}=\int_{{\mathbb{R}_{0}}}\left(S_{0}e^{u_{\beta}(z)}-K\right)^{+}\nu(dz)=\int_{{\mathbb{R}_{0}}}\left(S_{0}\beta e^{z}-(\beta-1)S_{0}-K\right)^{+}\nu(dz),
	\end{align}
    under the convention that $u_{\beta}(z)=-\infty$ for $z\notin A$. The previous expression is actually intuitive in light of the formula (\ref{eq:OTMOptPriceLETF}) and the dynamics of $Y^{o}$ as given in \eqref{eq:SDEYo}. Concretely, the leading order term of the call option price is determined only by the ``big'' jump component of the process as if $Y^{o}$ were simply a compound Poisson process $Y_{t}^{o}(x,x)=x+\int_{0}^{t}\int_{\mathbb{R}_{0}}u_{\beta}(z)\,\widetilde{M}_{A}^{\varepsilon}(ds,dz)$.
\item There is another way to interpret the approximation formula stated in Theorem \ref{thm:SmallTimeOTMCallOptLETF}. For $\beta\geq 1$, \eqref{eq:1stCoefOTMCallOptLETF1} implies that, in short-time, the price of an OTM European call on the LETF can ``closely'' be approximated by the price of an OTM European call on the underlying ETF, but with initial price $S_{0}\beta$ and modified strike $K_{\beta}=K+(\beta-1)S_{0}>0$. By contrast, for $\beta\leq 1-K/S_{0}<0$, \eqref{eq:1stCoefOTMCallOptLETF2} means that, in short time, the price of an OTM European call on the LETF is close to that of an OTM European put on the underlying ETF with initial price $|\beta|S_{0}$ and modified strike $K_{\beta}=-K-(\beta-1)S_{0}>0$. These observations in turn suggest a method to hedge OTM options on LETF near expiration using OTM options on ETF.
\item It is not hard to see that
    \begin{align}
    \frac{\partial b_{1}(x,K,\beta)}{\partial\beta}&=e^{x}\int_{\ln((Ke^{-x}-1+\beta)/\beta)}^{\infty}\left(e^{z}-1\right)h(z)\,dz>0,\quad\beta\geq 1,\\
    \frac{\partial b_{1}(x,K,\beta)}{\partial \beta}&=e^{x}\int_{-\infty}^{\ln((Ke^{-x}-1+\beta)/\beta)}\left(e^{z}-1\right)h(z)\,dz<0,\quad\beta\leq -1.
    \end{align}
	Thus, in short time, the call option price is increasing on $\beta\in[1,\infty)$, but decreasing on $\beta\in(-\infty,-1]$.
\item For $\beta\leq -1$, the extra condition that $\beta<1-Ke^{-x}$ ensures that the leading order of the short-time behavior of $\Pi(t;x,K,\beta)$ is of order $t$. Indeed, by \eqref{eq:1stCoefOTMCallOptLETFGen}, with ${L_{0}=S_{0}=e^{x}<K}$, the integrand does not vanish when $\beta e^{z}>Ke^{-x}+\beta-1$, which would never occur if $\beta\leq -1<0$ and $Ke^{-x}+\beta-1\geq 0$. By contrast, if $Ke^{-x}\geq 2$ and $\beta\in[1-Ke^{-x},-1]$, the first-order coefficient $b_{1}$ vanishes, and we have
    \begin{align*}
    \Pi(t;x,K,\beta)={o(t)},\quad t\rightarrow 0.
    \end{align*}
    There is another more intuitive interpretation for the above issue. By Remark \ref{rem:JumpSupport}, when $\beta\leq -1$, the sizes of jumps of the log-LETF $Y(x)$ are limited by $\ln(1-\beta)$. Hence, ignoring the diffusion part, the largest value $L$ can be after one jump is $e^{x+\ln(1-\beta)}=e^{x}(1-\beta)$. Hence, if $K>e^{x}(1-\beta)$, then it would require at least two (``big'') jumps to get there, which suggests a $O(t^{2})$ leading-order for $\Pi(t;x,K,\beta)$ as $t\rightarrow 0$. As it would be shown below, no extra condition of this type is needed for OTM put and ITM call options (i.e., $K<e^{x}$), since in this case $\beta\leq -1$ simply implies $\beta<1-Ke^{-x}$.
\end{itemize}
\end{remark}
\begin{remark}
\label{rmk:error}
Let us briefly comment on the error term of the approximation \eqref{eq:1stOrderAsyOTMCallOptLETF}. The proof of Theorem \ref{thm:SmallTimeOTMCallOptLETF} allows us to track down the different sources of errors: one for each term of the decomposition \eqref{eq:DecompPi}. In particular, we can further conclude that $o(t)$ is $O(t^{3/2})$ and that this error arises from the the term $I_{2}$ since all other terms therein give rise to $O(t^{2})$ errors. Concretely, Lemma \ref{lem:SmallTimeEstI1} shows that, for $\varepsilon>0$ and $t>0$ small enough,
\begin{align}
I_{1}(t)\leq 4K\exp\left(\frac{\left\|\gamma_{\varepsilon}\right\|_{\infty}}{2e\beta_{\varepsilon}}+\frac{4}{e^{2}\beta_{\varepsilon}^{2}}\left(\beta^{2}\|\sigma^{2}\|_{\infty}+\int_{\{|z|\leq\varepsilon\}} u_{\beta}^{2}(z)\,\nu(dz)\right)\right)t^{2},
\end{align}
where $\beta_{\varepsilon}=\ln(|\beta|(e^{\varepsilon}-1)+1)$, and $\gamma_{\varepsilon}$ is defined as in \eqref{Eq:Dfngammaepsilon}. For the term of $I_{2}(t)$, we have that
\begin{align}
\left|\frac{I_{2}(t)}{t}-b_{1}\right|\leq\lambda_{\varepsilon}(A)\widehat{C}_{1}t+\frac{3}{2}\widehat{C}_{2}\,\mathbb{E}\left(e^{u_{\beta}(J)}\right)\frac{e^{x}}{K}\widehat{C}_{3}\sqrt{t},
\end{align}
where $\widehat{C}_{1}$, $\widehat{C}_{2}$ and $\widehat{C}_{3}$ depend on the parameters of the model. More specifically, the constant $\widehat{C}_{1}$ is set to be an upper bound on
\begin{align}
\sup_{s\in[0,t]}\int_{K}^{\infty}\mathbb{E}\left(L_{\varepsilon}H_{0}\left(X_{s}^{\varepsilon}(x),Y_{s}^{\varepsilon}(x,x);z,\varepsilon,\beta\right)\right)dz,
\end{align}
with $H_{0}$ given as in \eqref{eq:FuntH0}. Therefore, $\widehat{C}_{1}$ can be taken as
\begin{align}
\widehat{C}_{1}:=e^{x}\widehat{C}_{3}\|\gamma_{\varepsilon}\|_{\infty}\!\int_{\mathbb{R}}g_{U}(w;\varepsilon,\beta)e^{w}\,dw+\frac{1}{2}e^{x}\widehat{C}_{3}\left(\beta^{2}\|\sigma^{2}\|_{\infty}\!+\!\int_{\{|z|\leq\varepsilon\}}\!\!e^{|u_{\beta}(z)|}u_{\beta}^{2}(z)h(z)\,dz\right)\int_{\mathbb{R}}\left|g'_{U}(w)\right|e^{w}\,dw,
\end{align}
where $g_{U}(\,\cdot\,;\varepsilon,\beta)$ denotes the density of $U_{\varepsilon}:=u_{\beta}(J_{\varepsilon})$ and $\widehat{C}_{3}$ is defined as below. The constant $\widehat{C}_{2}$ can be deduced from \eqref{Eq:BndC2} and can be set as
\begin{align}
\widehat{C}_{2}&:=c^{1/2}+\nu(A^c)+|\beta|\int_{A\cap\{|z|\geq\varepsilon\}}\left|e^{z}-1\right|h(z)\,dz,
\end{align}
with $c=c(\varepsilon,\beta,\|\sigma\|)$ given as in \eqref{eq:Constc}. The constant $\widehat{C}_{3}$ is set to be a bound for $\mathbb{E}(e^{Y_{1}^{\varepsilon}(x,0)})$. As can be seen from \eqref{Eq:DfnC4} and the argument thereafter, $\widehat{C}_{3}$ can be taken as
\begin{align}
\widehat{C}_{3}&:=1\vee\exp\left(\nu(A^c)-\beta\int_{A\cap\{|z|\geq\varepsilon\}}\left(e^{z}-1\right)h(z)\,dz\right).
\end{align}
Finally, we have that
\begin{align}
I_{3}(t)\leq\widehat{C}_{4}\,\lambda_{\varepsilon}^{2}(A)\,e^{x}\,\frac{t^{2}}{1-\lambda_{\varepsilon}(A)t},
\end{align}
where, as shown in the proof of Lemma \ref{lem:SmallTimeEstI3}, $\widehat{C}_{4}$ is set as a constant such that
\begin{align}
\frac{1}{n!}\widehat{D}^{n+1}\left(\frac{1}{e^{1/\sqrt{n}}-1}\right)^{n+1}\leq\widehat{C}_{4},\quad\text{for all }\,n\in\mathbb{N},
\end{align}
where $\widehat{D}:=3\Lambda_{\varepsilon/2}\exp(3\|\gamma_{\varepsilon}\|_{\infty}+(9/2)\widetilde{D}(1+e^{3\beta_{\varepsilon}}))$.
\end{remark}

Next, we study the small-time asymptotic behavior of an OTM European put option on the LETF $L$, with maturity $t>0$ and strike price $K<e^{x}$. As
above, we denote $\Theta(t;x,K,\beta)$ the time-zero price of the OTM put option. Then,
\begin{align}
\Theta(t;x,K,\beta)=\mathbb{E}\left(\left(K-L_{t}\right)^{+}\right)&=\mathbb{E}\left({\bf
1}_{\{\tau>t\}}\left(K-e^{Y_{t}}\right)^{+}\right)+K\,\mathbb{P}\left(\tau\leq t\right)\\
&=e^{-t\nu(A^{c})}\mathbb{E}\left(\left(K-e^{Y_{t}^{o}(x,x)}\right)^{+}\right)+K\,\mathbb{P}\left(\tau\leq t\right).
\end{align}
From the definition of the default time \eqref{eq:DefaultTimeLETF},
\begin{align}
\mathbb{P}\left(\tau\leq t\right)=\mathbb{P}\left(N\left([0,t]\times A^{c}\right)\geq 1\right)=1-e^{-t\nu(A^{c})}.
\end{align}
It remains to study the first term above, hereafter denoted by $\widetilde{\Theta}(t;x,K,\beta)$. Similar to \eqref{eq:DecompPi} - \eqref{eq:I3}, we can decompose $\widetilde{\Theta}(t;x,K,\beta)$ by conditioning on the number of ``big'' jumps occurring up to time $t$:
\begin{align}
\widetilde{\Theta}(t;x,K,\beta)=e^{-t\nu(A^{c})}e^{-t\lambda_{\varepsilon}(A)}\left(\widetilde{I}_{1}(t)+\widetilde{I}_{2}(t)+\widetilde{I}_{3}(t)\right),
\end{align}
where
\begin{align}
\widetilde{I}_{1}(t)=\widetilde{I}_{1}(t;x,K,\varepsilon,\beta)&:=\mathbb{E}\left(\left.\left(K-e^{Y^{o}_{t}(x,x)}\right)^{+}\,\right|N_{t}^{\varepsilon}(A)=0\right)=\mathbb{E}\left(\left(K-e^{Y^{\varepsilon}_{t}(x,x)}\right)^{+}\right),\\
\widetilde{I}_{2}(t)=\widetilde{I}_{2}(t;x,K,\varepsilon,\beta)&:=t\lambda_{\varepsilon}(A)\,\mathbb{E}\left(\left.\left(K-e^{Y^{o}_{t}(x,x)}\right)^{+}\,\right|N_{t}^{\varepsilon}(A)=1\right)\\
\label{eq:I2b} &\,\,=K\lambda_{\varepsilon}(A)\int_{0}^{t}\mathbb{E}\left(\widetilde{G}_{t-s}\left(X_{s}^{\varepsilon}(x),Y_{s}^{\varepsilon}(x,x-\ln
K);\varepsilon\right)\right)ds,\\
\widetilde{I}_{3}(t)=\widetilde{I}_{3}(t;x,K,\varepsilon,\beta)&:=t^{2}\lambda_{\varepsilon}^{2}(A)\sum_{n=2}^{\infty}\frac{\left(\lambda_{\varepsilon}(A)\,t\right)^{n-2}}{n!}\,\mathbb{E}\left(\left.\left(K-e^{Y^{o}_{t}(x,x)}\right)^{+}\,\right|N_{t}^{\varepsilon}(A)=n\right).
\end{align}
Above, we have set
\begin{align}
\widetilde{G}_{t}\left(\bar{x},\bar{y};x,z,\varepsilon\right):=\mathbb{E}\left(\left(1-e^{Y^{\varepsilon}_{t}(\bar{x}+J_{\varepsilon},\bar{y}+u_{\beta}(J_{\varepsilon}))}\right)^{+}\right),\quad
t\geq 0,\quad\bar{x},\bar{y}\in\mathbb{R},
\end{align}
where, again, $J_{\varepsilon}$ is a random variable, independent of $X^{\varepsilon}$ and $Y^{\varepsilon}$, with density \eqref{eq:DenJumpZ2epsA}. By Lemma \ref{lem:SmallTimeEstI3} as well as the following formula (where $Z$ represents any random variable)
\begin{align}
\mathbb{E}\left((K-Z){\bf 1}_{\{Z<K\}}\right)=\int_{0}^{K}\mathbb{P}(Z<z)\,dz,
\end{align}
it is easy to see that $\widetilde{I}_{3}(t)=O(t^{2})$, for all $t\in[0,1]$. Moreover, the analysis of the small-time asymptotic behavior of $\widetilde{I}_{1}(t)$, $\widetilde{I}_{2}(t)$ and $\widetilde{\Theta}(t;x,K,\beta)$ is very similar to those of $I_{1}(t)$, $I_{2}(t)$ and $\Pi(t;x,K,\beta)$ presented in Lemma \ref{lem:SmallTimeEstI1}, Lemma \ref{lem:SmallTimeExpG}, Lemma \ref{lem:SmoothH0} and Theorem \ref{thm:SmallTimeOTMCallOptLETF}. Below, we will only present the results while skipping all proofs.
\begin{lemma}\label{lem:SmallTimeEstTildeI1}
Let Assumption \ref{assump:UnifBdVol} and Assumption \ref{assump:SmthBdLevyDen} be valid. Let $K<e^{x}$ and let $\beta\in(-\infty,-1]\cup[1,\infty)$. Then for any $n\in\mathbb{N}$, and any $\varepsilon\in(0,{\ln((e^{(x-\ln K)/(2n)}-1)/|\beta|+1)}\wedge |\ln(1-\beta^{-1})|\wedge 1)$, there exists $\widetilde{C}_{2}>0$, depending on $K$, $x$, $\varepsilon$, $|\beta|$ and $\|\sigma\|_{\infty}$, such that $|\widetilde{I}_{1}(t)|\leq\widetilde{C}_{2}\,t^{n}$, for all $t\in[0,1]$.
\end{lemma}
\begin{lemma}\label{lem:SmallTimeExpTildeG}
Let Assumption \ref{assump:UnifBdVol} and Assumption \ref{assump:SmthBdLevyDen} be valid. Let $\beta\in(-\infty,-1]\cup[1,\infty)$ and let
$\varepsilon\in(0,|\ln(1-1/\beta)|\wedge 1)$. Then, there exists a constant $\widetilde{C}_{3}>0$, depending only on $\beta$, $\varepsilon$ and
$\|\sigma\|_{\infty}$, such that for any $\bar{x},\bar{y}\in\mathbb{R}$, and any $t\geq 0$,
\begin{align}
\left|\widetilde{G}_{t}(\bar{x},\bar{y};\varepsilon,\beta)-{\widetilde{G}_{0}(\bar{x},\bar{y};\varepsilon,\beta)}\right|\leq\widetilde{C}_{3}\,\mathbb{E}\left(e^{u_{\beta}(J_{\varepsilon})}\right)e^{\bar{y}}\sqrt{t}.
\end{align}
\end{lemma}
\begin{lemma}\label{lem:SmoothH0}
Let Assumption \ref{assump:SmthBdLevyDen} be valid. Let $\beta\in(-\infty,-1]\cup[1,\infty)$ and let $\varepsilon\in(0,|\ln(1-1/\beta)|\wedge 1)$. For any fixed $z\geq 1$, let
\begin{align}\label{eq:FuntTildeH0}
\widetilde{H}_{0}(\bar{y};z,\varepsilon,\beta):=\mathbb{P}\left(u_{\beta}(J)<\ln z-\bar{y}\right).
\end{align}
Then, for $\beta\geq 1$, $\widetilde{H}_{0}(\,\cdot\,;z,\varepsilon,\beta)\in C_{b}^{2}(\mathbb{R})$. For $\beta\leq -1$, $\widetilde{H}_{0}(\,\cdot\,;z,\varepsilon,\beta)\in C_{b}^{2}(\mathbb{R})$, provided that the L\'{e}vy density $h$ satisfies \eqref{eq:AdContLevyDen}.
\end{lemma}
\begin{theorem}\label{thm:SmallTimeOTMPutOptLETF}
Let $\Theta(t;x,K,\beta)$ be the time-zero price of a European put option on the LETF $L$ (with leverage ratio $\beta$), where $L_{0}=e^{x}$, with strike price $K$ and maturity $t$. Let $K<e^{x}$, and let $\beta\in(-\infty,-1]\cup[1,\infty)$. Suppose Assumption \ref{assump:UnifBdVol} and Assumption \ref{assump:SmthBdLevyDen} are valid. For $\beta\geq 1$, we have
\begin{align}\label{eq:1stOrderAsyOTMPutOptLETF}
\Theta(t;x,K,\beta)=\widetilde{b}_{1}\,t+o(t),\quad t\rightarrow 0,
\end{align}
where
\begin{align}\label{eq:1stCoefOTMPutOptLETF1}
\widetilde{b}_{1}=\widetilde{b}_{1}(x,K,\beta):=K\nu(A^{c})+\int_{\ln(1-\beta^{-1})}^{\ln((Ke^{-x}-1+\beta)/\beta)}\left[K-\beta
e^{x+z}+(\beta-1)e^{x}\right]h(z)\,dz.
\end{align}
When $\beta\leq -1$, \eqref{eq:1stOrderAsyOTMPutOptLETF} remains valid under the additional condition \eqref{eq:AdContLevyDen}, with
\begin{align}\label{eq:1stCoefOTMPutOptLETF2}
\widetilde{b}_{1}=\widetilde{b}_{1}(x,K,\beta):=K\nu(A^{c})+\int_{\ln((Ke^{-x}-1+\beta)/\beta)}^{\ln(1-\beta^{-1})}\left[K-\beta e^{x+z}+(\beta-1)e^{x}\right]h(z)\,dz.
\end{align}
\end{theorem}

Using put-call parity, we can also obtain the following result for the corresponding ITM European call option on the LETF.
\begin{corollary}\label{cor:SmallTimeITMCallOptLETF}
Let $\Pi(t;x,K,\beta)$ be the time-zero price of a European call option on the LETF, where $L_{0}=e^{x}$, with strike price $K$ and maturity $t$. Under conditions of Theorem \ref{thm:SmallTimeOTMPutOptLETF}, we have
\begin{align}
\Pi(t;x,K,\beta)=e^{x}-K+\widetilde{b}_{1}t+o(t),\quad t\rightarrow 0,
\end{align}
where $\widetilde{b}_{1}=\widetilde{b}_{1}(x,K,\beta)$ is given by \eqref{eq:1stCoefOTMPutOptLETF1} and \eqref{eq:1stCoefOTMPutOptLETF2},
respectively, for the case when $\beta\geq 1$ and $\beta\leq -1$.
\end{corollary}

\begin{remark}
As with European calls on LETFs, we can write \eqref{eq:1stOrderAsyOTMPutOptLETF} - \eqref{eq:1stCoefOTMPutOptLETF2} in the more appealing and unified form:
\begin{align}\label{eq:1stCoefOTMPutOptLETFGen}
\widetilde{b}_{1}=\int_{\mathbb{R}_{0}}\left(K-{S_{0}}e^{u_{\beta}(z)}\right)^{+}\nu(dz)=\int_{\mathbb{R}_{0}}\left(K-S_{0}\beta e^{z}+(\beta-1)S_{0}\right)^{+}\nu(dz),
\end{align}
under the convention that $u_{\beta}(z)=-\infty$ for $z\notin A$. Therefore, for $\beta\geq 1$, \eqref{eq:1stOrderAsyOTMPutOptLETF} implies that, in short-time, the price of an OTM European put on the LETF can ``closely'' be approximated by the price of an OTM European put on the underlying ETF, but with initial spot price $S_{0}\beta$ and modified strike $K_{\beta}=K+(\beta-1)S_{0}>0$. By contrast, for $\beta\leq -1$, \eqref{eq:1stOrderAsyOTMPutOptLETF} means that, in short time, the price of an OTM European put on the LETF is close to that of an OTM European call on the underlying ETF with initial price $|\beta|S_{0}$ and modified strike $K_{\beta}=(1-\beta)S_{0}-K>0$. Again, these observations in turn suggest a method to hedge OTM options on LETF near expiration using OTM options on ETF.
\end{remark}
\begin{remark}
Our framework can be generalized to the case where the jump size of the log-ETF (and thus the jump size of the log-LETF) is state-dependent, e.g., $\theta(X_{t-},z)$. Although the predefault domain $A$ becomes random in this generalized model, similar short-time asymptotic behavior can be obtained for the off-the-money options under some additional regularity conditions on the function $\theta$. Another interesting extension of our results is to consider the small-time asymptotics of the at-the-money options under the current model. But this is out of the scope of the present article, and will be studied elsewhere.
\end{remark}

\section{The Implied Volatility}\label{sec:IVLETFs}

In this section, we will apply the small-time asymptotic results of not-at-the-money European call (equivalently, put) options on the LETF $L$, presented in Theorem \ref{thm:SmallTimeOTMCallOptLETF} and Corollary \ref{cor:SmallTimeITMCallOptLETF} (equivalently, Theorem
\ref{thm:SmallTimeOTMPutOptLETF} and Corollary \ref{cor:SmallTimeITMPutOptLETF}) above, to derive the small-time asymptotics for the
corresponding not-at-the-money Black-Scholes implied volatility. Throughout this section, let $C_{\text{BS}}(t;x,K,\sigma)$ be the price of the European call option on the ETF under the Black-Scholes model, with strike price $K$, maturity $t$, initial log-ETF price $x$, and constant volatility $\sigma$. Let $\hat{\sigma}(t)=\hat{\sigma}(t;x,K,\beta)$ be the corresponding Black-Scholes implied volatility of the call option price \eqref{eq:OTMOptPriceLETF0}, namely, $\hat{\sigma}(t)$ is such that $C_{\text{BS}}(t;x,K,\hat{\sigma}(t))=\Pi(t;x,K,\beta)$.

We first recall the following small-time asymptotic expansion of not-at-the-money Black-Scholes European-call option price (cf.~\cite[Corollary 3.4]{FordeJacquierLee:2012}, assuming zero interest rates): for fixed $\sigma,K>0$ and $x\in\mathbb{R}$ such that $K\neq e^{x}$, as $t\rightarrow 0$,
\begin{align}\label{eq:AsyExpBSOTMOpt}
C_{\text{BS}}(t;x,K,\sigma)=\left(e^{x}-K\right)^{+}+\frac{K\sigma^{3}t^{3/2}}{\sqrt{2\pi}(\ln K-x)^{2}}\exp\left(-\frac{(\ln
K-x)^{2}}{2\sigma^{2}t}-\frac{\ln K-x}{2}\right)+O\left(t^{5/2}\right).
\end{align}
The following result summarizes the small-time asymptotic behavior of $\hat{\sigma}(t)$, as $t\rightarrow 0$. The proof is similar to those given in~\cite[Theorem 2.3]{FigueroaLopezForde:2012} and~\cite[Lemma 5.1]{FigueroaLopezGongHoudre:2012} for OTM call options on exponential L\'{e}vy assets, and is thus deferred to the appendix.
\begin{theorem}\label{thm:SmallTimeIVLETF}
Let Assumption \ref{assump:UnifBdVol} and Assumption \ref{assump:SmthBdLevyDen} be valid.
\begin{itemize}
\item [(i)] Let $K>e^{x}$. Then, as $t\rightarrow 0$,
    \begin{align}\label{eq:2ndOrderAsyIV}
    \hat{\sigma}^{2}(t)=\sigma_{1}(t)\left(1+\sigma_{2}(t)+o\left(\frac{1}{\ln(1/t)}\right)\right),
    \end{align}
    where
    \begin{align}\label{eq:1stCoefIV}
    \sigma_{1}(t)&=\sigma_{1}(t;x,K)=\frac{(\ln K-x)^{2}}{2t\ln(1/t)},\\
    \label{eq:2ndCoefIV} \sigma_{2}(t)&=\sigma_{2}(t;x,K,\beta)=\frac{1}{\ln(1/t)}\ln\left(\frac{4\sqrt{\pi}\,b_{1}(x,K,\beta)e^{(\ln K-x)/2}}{K\left|\ln K-x\right|}\left(\ln\left(\frac{1}{t}\right)\right)^{3/2}\right),
    \end{align}
    and where $b_{1}(x,K,\beta)$ is given by \eqref{eq:1stCoefOTMCallOptLETF1} and \eqref{eq:1stCoefOTMCallOptLETF2}, respectively, when
    $\beta\geq 1$ and {$\beta\in (-\infty,-1]\cap(-\infty,1-Ke^{-x})$}.
\item [(ii)] Let $K<e^{x}$. Then, as $t\rightarrow 0$, \eqref{eq:2ndOrderAsyIV}, \eqref{eq:1stCoefIV} and \eqref{eq:2ndCoefIV} remain valid, with
    $b_{1}(x,K,\beta)$ in the expression \eqref{eq:2ndCoefIV} of $\sigma_{2}(t)$ replaced by $\widetilde{b}_{1}(x,K,\beta)$, given
    respectively by \eqref{eq:1stCoefOTMPutOptLETF1} and \eqref{eq:1stCoefOTMPutOptLETF2} when $\beta\geq 1$ and $\beta\leq -1$.
\end{itemize}
\end{theorem}
\begin{remark}
One may naturally wonder how the implied volatility smile of the leveraged product is related to that of the underlying ETF. This point has
received some attention in the literature, as stated in the introduction. We can similarly raise the same question here, at least for short-maturity options, via the formulas \eqref{eq:2ndOrderAsyIV}-\eqref{eq:2ndCoefIV}. These show that the leading term, $\hat\sigma_{1}$, is not
affected by the leverage $\beta$. However, the leverage already appears in the second order term $\hat\sigma_{2}$, so that, in small time,
\begin{align}
\hat\sigma^{2}(t,K;\beta)=\hat\sigma^{2}(t,K;1)+\hat\sigma_{1}(t)\ln\left(\frac{b_{1}(x,K,\beta)}{b_{1}(x,K,1)}\right)+\rm{h.o.t.},
\end{align}
where ${\rm h.o.t.}$ means ``higher order terms.'' Thus, in terms of the log-moneyness $\kappa=\ln(K/S_{0})$, the correction term depends on the ratio
\begin{align}
\frac{b_{1}(x,K,\beta)}{b_{1}(x,K,1)}=\frac{\int\left(\beta e^{z}-(\beta-1)-e^{\kappa}\right)^{+}\nu(dz)}{\int\left(e^{z}-e^{\kappa}\right)^{+}\nu(dz)}.
\end{align}
It is important to remark that our results in Theorems \ref{thm:SmallTimeOTMCallOptLETF} and \ref{thm:SmallTimeOTMPutOptLETF} together with the methodology in~\cite{GaoLee:2014} would allow us to derive expansions for the implied volatility with an error of order $O(|\ln\left(1/t\right)|^{-j})$ for arbitrarily large $j\geq 1$ (see~\cite[Section 8.2]{GaoLee:2014}). For simplicity, we just consider here the second-order expansion.
\end{remark}

\section{Numerical Examples}\label{sec:numerical}

In this section we provide two examples, which illustrate the numerical accuracy and flexibility of the implied volatility approximation given in
Theorem \ref{thm:SmallTimeIVLETF}.

\subsection{Kou Double Exponential Jumps With Local Volatility}\label{sec:kou}

In our first example, we consider a local volatility model with compound Poisson jumps (i.e., $\nu(\mathbb{R}_{0})<\infty$). Specifically, the local volatility function $\sigma$ and L\'{e}vy density $h$ are given by
\begin{align}\label{eq:local.vol}
\sigma(x)&=a+b\tanh cx, && a>|b|>0,\\
\label{eq:kou} h(z)&=\lambda\left(p {\bf 1}_{\{z>0\}}\eta_{1}e^{-\eta_{1}z}+q{\bf 1}_{\{z<0\}}\eta_{2}e^{\eta_{2}z}\right), && \lambda,\,q,\,p,\,\eta_{2}>0,\quad\eta_{1}>1,\quad p+q=1.
\end{align}
Note that the local volatility function is bounded: $a-|b|<\sigma(x)<a+|b|$ for all $x\in\mathbb{R}$. Also, if $bc<0$, then $\sigma$ is decreasing, which is consistent with the leverage effect. The L\'{e}vy measure $\nu$ in \eqref{eq:kou} first appeared in a financial context in~\cite{Kou:2002}. The net jump intensity is $\lambda$. When a jump occurs, it is positive with probability $p$. The positive jumps (respectively, absolute values of negative jumps) are exponentially distributed with parameter $\eta_{1}$ (respectively, parameter $\eta_{2}$).

\begin{figure}
\centering
\begin{tabular}{cc}
\includegraphics[width=0.45\textwidth]{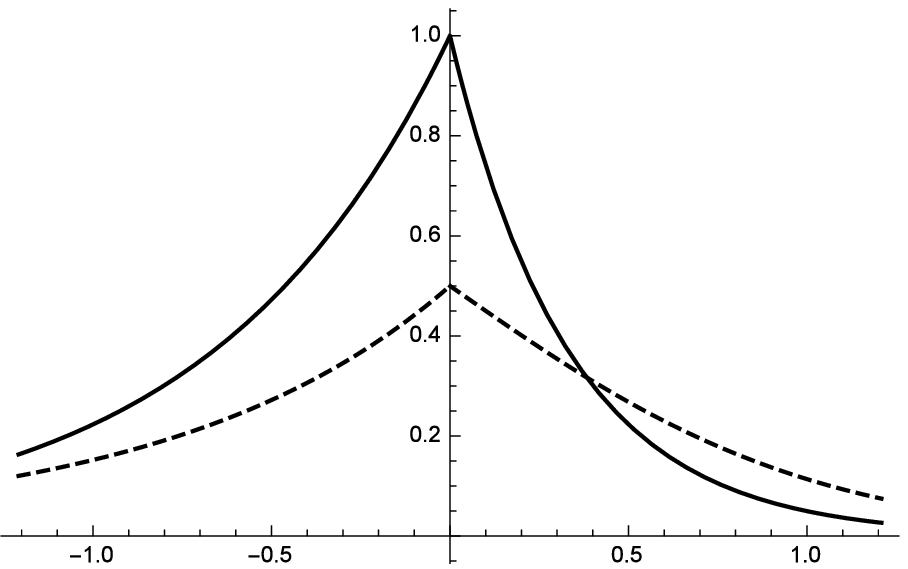} & \includegraphics[width=0.45\textwidth]{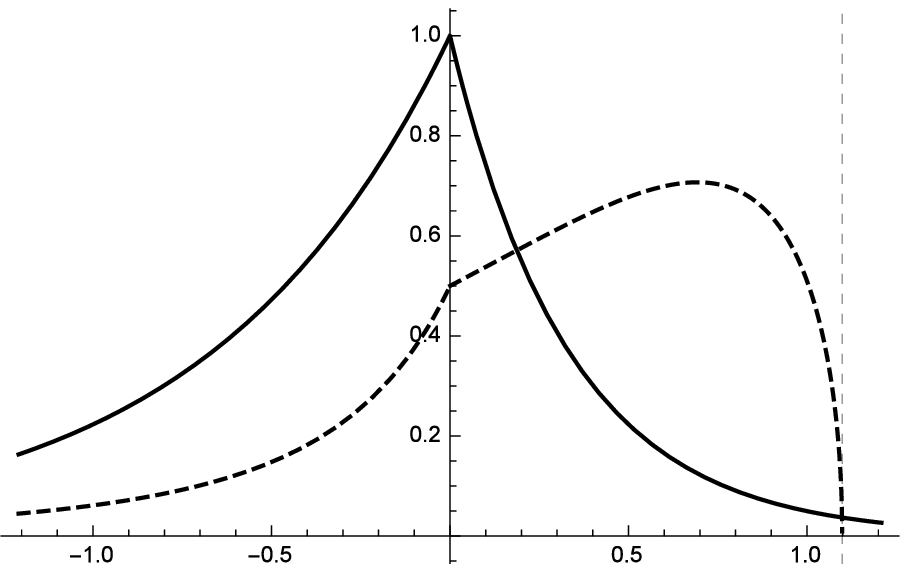} \\
$\beta=+2$ & $\beta=-2$
\end{tabular}
\caption{
We compare the L\'{e}vy density $h$ of the log-ETF $X$ (solid line) to the L\'{e}vy density $g$ of the log-LETF $Y$ (dashed line) with $h$ given by
\eqref{eq:kou}. For this plot we fix $\lambda=1$, $\eta_{1}=3$, $\eta_{2}=3/2$, $p=1/3$ and $q=2/3$. These parameters are chosen purely to illustrate the effect of $\beta$. The vertical dashed line on the right indicates the upper limit of the support of $g$.
}
\label{fig:kou.density}
\end{figure}

It is interesting to observe how the leverage ratio $\beta$ affects the L\'{e}vy density of the log-LETF $Y$. Recall that $Y$ has a L\'{e}vy measure given by $\nu\circ u_{\beta}^{-1}:=\pi$. Thus, denoting by $g$ the density of $\pi$ we have
\begin{align}\label{eq:levy.density}
g(z)&=h\left(u_{\beta}^{-1}(z)\right)\left(u_{\beta}^{-1}\right)'(z).
\end{align}
In Figure \ref{fig:kou.density} we plot $h$ and $g$ for various values of $\beta$ when $h$ is given by \eqref{eq:kou}. Note that, when $\beta\leq
-1$, the support of $g$ is $(-\infty,\ln(1-\beta))$.

\begin{figure}
\centering
\begin{tabular}{cc}
\includegraphics[width=0.45\textwidth]{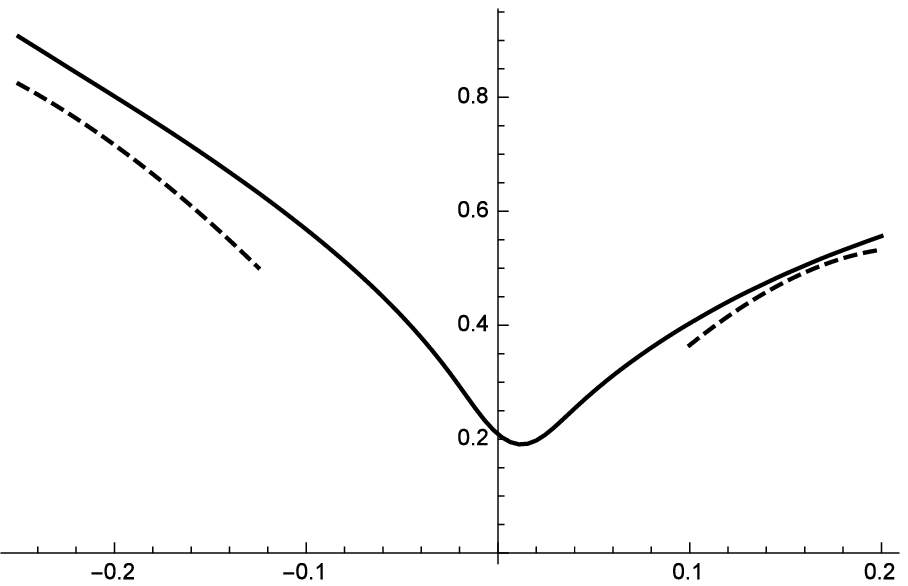} & \includegraphics[width=0.45\textwidth]{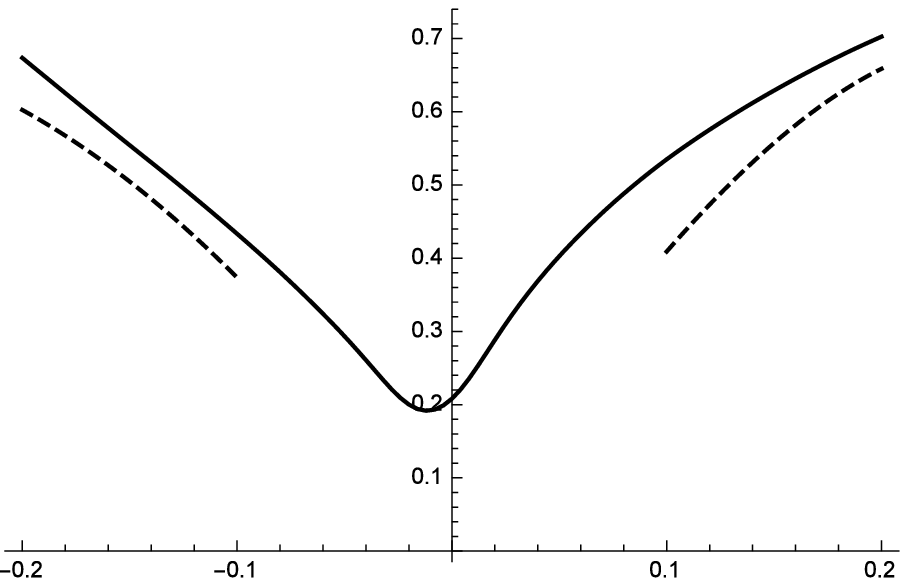} \\
$\beta=+1$ & $\beta=-1$\\[1em]
\includegraphics[width=0.45\textwidth]{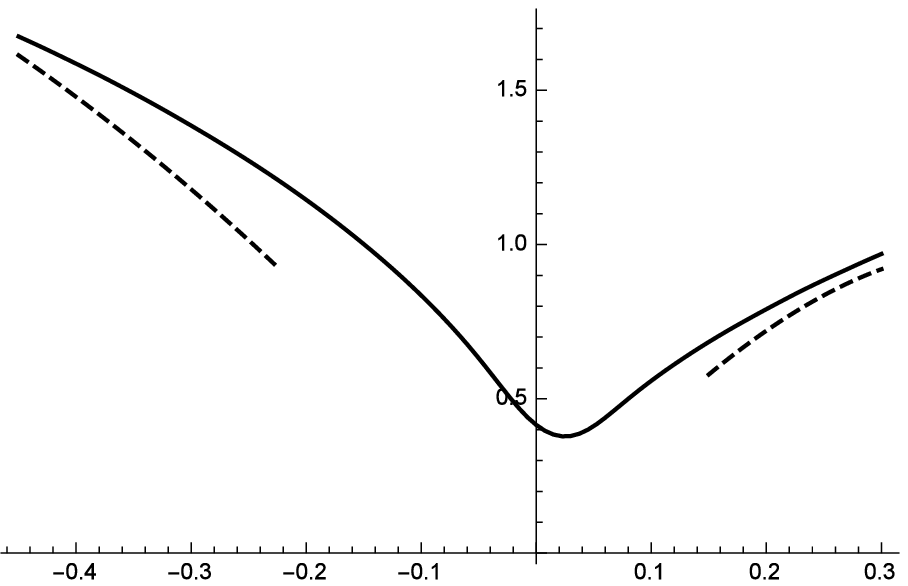} & \includegraphics[width=0.45\textwidth]{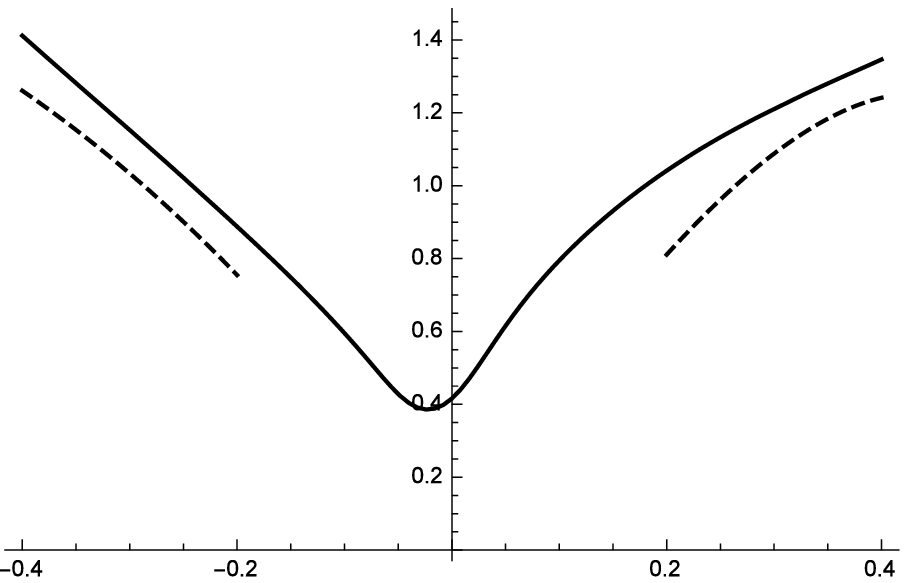} \\
$\beta=+2$ & $\beta=-2$
\end{tabular}
\caption{
Here we plot implied volatility as a function of log-moneyness $\ln K-x$ for the local volatility model with double exponential jumps discussed
in Section \ref{sec:kou}. The solid line represents implied volatilities computed via Monte Carlo simulation. The dashed lines indicate implied
volatilities computed using the approximation given in Theorem \ref{thm:SmallTimeIVLETF}.
}
\label{fig:kou.impvol}
\end{figure}

To illustrate the accuracy of our implied volatility expansion, we fix the following parameters
\begin{align}
x & =0, & t & =5/365, & a & =0.05, & b & =-0.02, & c & =0.5,\\
\lambda & =15, & p & =1/3, & q & =2/3, & \eta_{1} & =25, & \eta_{2} & =15.
\end{align}
The parameters for $\nu$ are in line with the range of values considered in~\cite{KouWang:2004}. We compute prices of call options on $L$ via Monte Carlo simulation using a standard Euler scheme. We fix a time-step of $t/100$ and run $1,000,000$ sample paths. Option prices are converted to implied volatilities by inverting the Black-Scholes formula numerically. In Figure \ref{fig:kou.impvol} we plot the implied volatilities resulting from the Monte Carlo simulation along with the approximation implied volatilities computed via Theorem \ref{thm:SmallTimeIVLETF}. Figure \ref{fig:kou.impvol} shows that, for $\beta\in\{-2,-1,+1,+2\}$, the implied volatility approximation closely matches the slope of the true implied volatility. However, the former falls below the latter at all strikes.

\subsection{Variance Gamma {Jumps with Local Volatility}}\label{sec:vg}

In this example, we consider a local volatility model with infinite activity jumps (i.e., $\nu(\mathbb{R}_{0})=\infty$). Specifically, the local volatility function $\sigma$ is given by \eqref{eq:local.vol} and L\'{e}vy density $h$ is the \emph{variance gamma density}
\begin{align}\label{eq:vg}
h(z)&=\frac{1}{\kappa |z|}\exp\left(Az-B|z|\right), & A &=\frac{\theta}{\sigma^{2}}, & B &=\sqrt{A^{2}+\frac{2}{\kappa\sig^{2}}} , && \kappa,\sigma>0,
\end{align}
which first appeared in finance in~\cite{MadanCarrChang:1998}. The L\'{e}vy density $h$ corresponds to the L\'{e}vy density of drifted Brownian motion $\sigma W_{t}+\theta t$, which is time-changed by a Gamma subordinator with parameter $\kappa$.

\begin{figure}
\centering
\begin{tabular}{cc}
\includegraphics[width=0.45\textwidth]{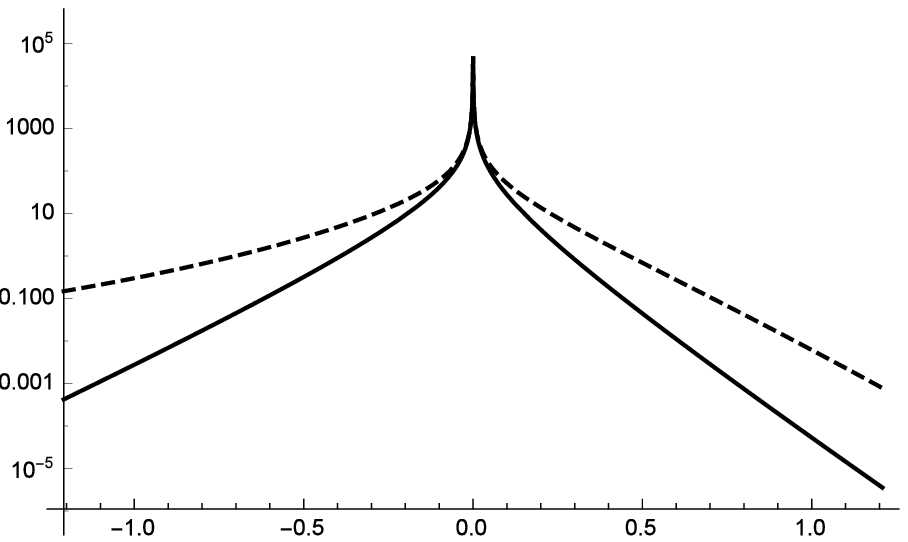} & \includegraphics[width=0.45\textwidth]{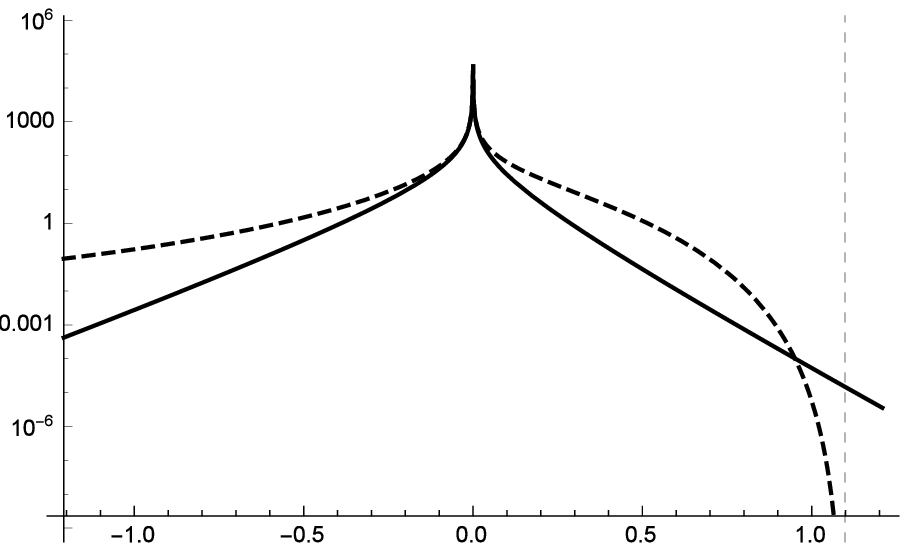} \\
$\beta=+2$ & $\beta=-2$
\end{tabular}
\caption{
We compare the L\'{e}vy density $h$ of the log-ETF $X$ (solid line) to the L\'{e}vy density $g$ of the log-LETF $Y$ (dashed line) with $h$ given by
\eqref{eq:vg}. For this plot we use the parameters given in \eqref{eq:vg.params}. The vertical dashed line on the right indicates the upper limit of
the support of $g$.
}
\label{fig:vg.density}
\end{figure}

In order to test the accuracy of our implied volatility expansion, we fix the following parameters
\begin{align}
x & =0, & t & =5/365, & a & =0.005, & b & =-0.002,\\
\label{eq:vg.params} c & =0.5, & \kappa & =0.1083, & \theta & =-0.3726, & \sigma & =0.4344.
\end{align}
The L\'{e}vy density parameters and initial level of volatility are those obtained in~\cite{CarrGemanMadanYor:2002} by calibrating the Variance Gamma model to IBM closing option prices on February 10th, 1999 with maturities of 1 and 2 months. In Figure \ref{fig:vg.density} we plot $h$ and $g$ for various values of $\beta$ when $h$ is given by \eqref{eq:vg}. Because the densities $h$ and $g$ blow up at the origin, we use a $\ln$ scale on the vertical axis.

\begin{figure}
\centering
\begin{tabular}{cc}
\includegraphics[width=0.45\textwidth]{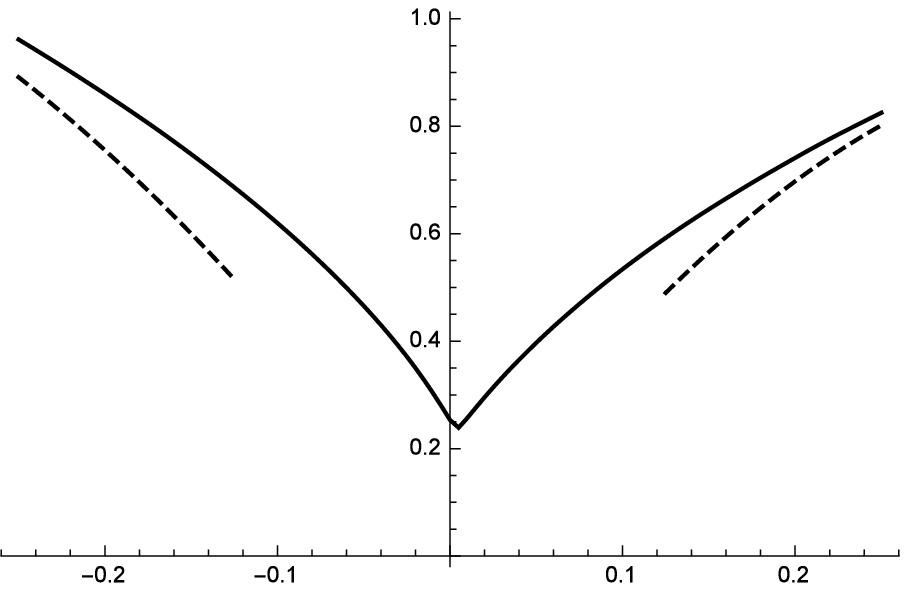} & \includegraphics[width=0.45\textwidth]{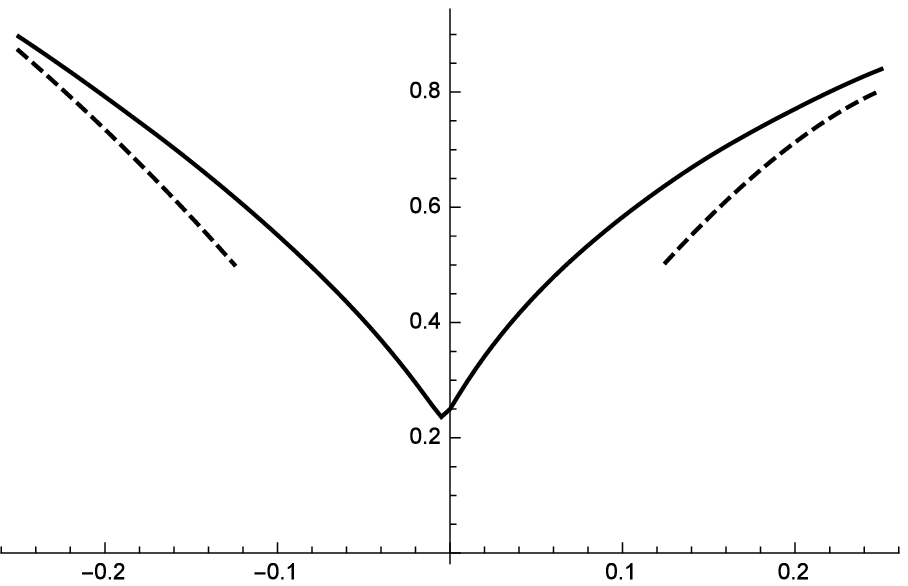} \\
$\beta=+1$ & $\beta=-1$\\[1em]
\includegraphics[width=0.45\textwidth]{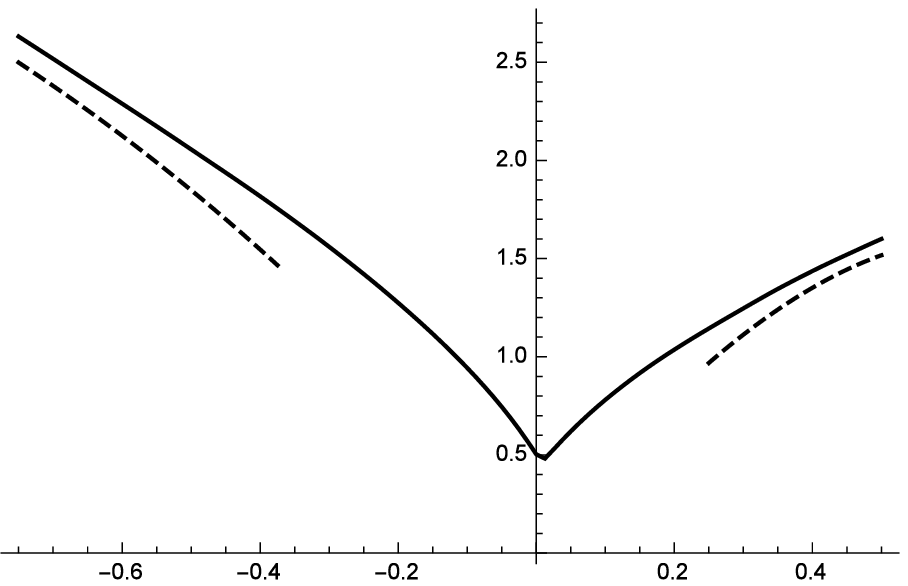} & \includegraphics[width=0.45\textwidth]{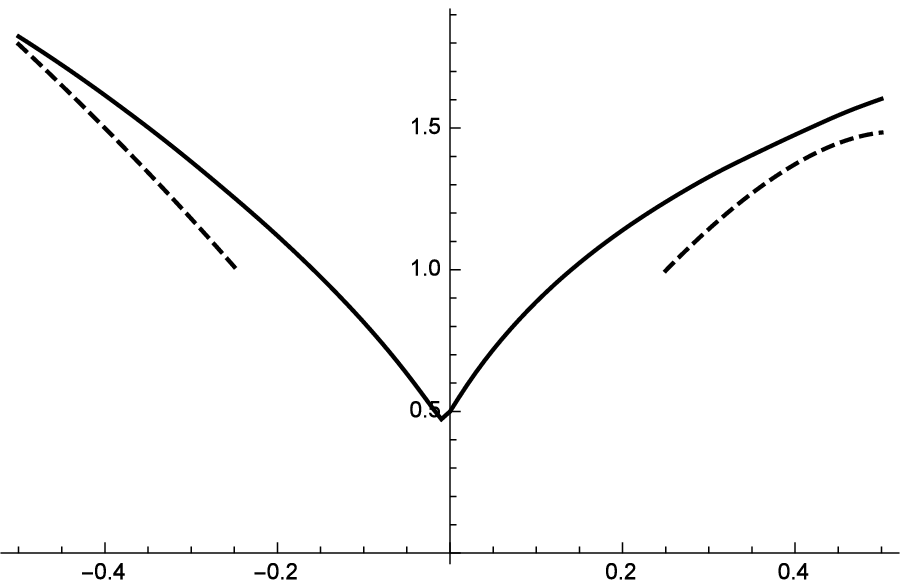} \\
$\beta=+2$ & $\beta=-2$
\end{tabular}
\caption{
Here we plot implied volatility as a function of log-moneyness $\ln K-x$ for the local volatility model with variance gamma jumps discussed in Section \ref{sec:vg}. The solid line represents implied volatilities computed via Monte Carlo simulation. The dashed lines indicate implied volatilities computed using the approximation given in Theorem \ref{thm:SmallTimeIVLETF}.
}
\label{fig:vg.impvol}
\end{figure}

We compute prices of call options on $L$ via Monte Carlo simulation using a standard Euler scheme. We fix a time-step of $t/100$ and run $1,000,000$ sample paths. Note that increments of the Variance Gamma process can be simulated exactly on a fixed time grid using Algorithm 6.11 in~\cite{ContTankov:2004}. Option prices are converted to implied volatilities by inverting the Black-Scholes formula numerically. In Figure \ref{fig:vg.impvol} we plot the implied volatilities resulting from the Monte Carlo simulation along with the approximation implied volatilities computed via Theorem \ref{thm:SmallTimeIVLETF}. Similar to the Kou Double Exponential case, Figure \ref{fig:kou.impvol} shows that, for $\beta\in\{-2,-1,+1,+2\}$, the implied volatility approximation closely matches the slope of the true implied volatility. However, the former falls below the latter at all strikes.

\appendix

\section{Additional Proofs}

\noindent
\textbf{Proof of Lemma \ref{lem:Dynkin}.} Applying It\^{o}'s formula, we have
\begin{align}
f\!\left(X_{t}^{\varepsilon}(x),Y_{t}^{\varepsilon}(x,y)\right)&=f(x,y)+\int_{0}^{t}L_{\varepsilon}f\!\left(X_{s}^{\varepsilon}(x),Y_{s}^{\varepsilon}(x,y)\right)ds\\
&\quad\,+\int_{0}^{t}\left(\frac{\partial f}{\partial x}\!\left(X_{s}^{\varepsilon}(x),Y_{s}^{\varepsilon}(x,y)\right)+\beta\frac{\partial
f}{\partial
y}\!\left(X_{s}^{\varepsilon}(x),Y_{s}^{\varepsilon}(x,y)\right)\right)\sigma\!\left(X_{s}^{\varepsilon}(x)\right)d\widetilde{W}_{s}\\
\label{eq:ItoXYeps}
&\quad\,+\int_{0}^{t}\!\int_{\mathbb{R}_{0}}\left(f\left(X_{s}^{\varepsilon}(x)\!+\!z,Y_{s}^{\varepsilon}(x,y)\!+\!u_{\beta}(z)\right)\!-\!f\left(X_{s}^{\varepsilon}(x),Y_{s}^{\varepsilon}(x,y)\right)\right)\widetilde{M}_{A}^{\varepsilon,1}(ds,dz).
\end{align}
Above, the Brownian integral is an $\mathbb{F}$-martingale under $\mathbb{P}$ due to Assumption \ref{assump:UnifBdVol} and since $f\in
C_{b}^{2}(\mathbb{R}^{2})$. Also, due to Assumption \ref{assump:UnifBdVol}, the stochastic integral with respect to $\widetilde{M}_{A}^{\varepsilon,1}$ is an $\mathbb{F}$-martingale under $\mathbb{P}$ since
\begin{align}
&\int_{0}^{t}\int_{0<|z|<1}\left(f\left(X_{s}^{\varepsilon}(x)+z,Y_{s}^{\varepsilon}(x,y)+u_{\beta}(z)\right)-f\left(X_{s}^{\varepsilon}(x),Y_{s}^{\varepsilon}(x,y)\right)\right)^{2}h_{A}^{\varepsilon,1}(z)\,dz\,ds\\
&\quad\,\leq 2\int_{0}^{t}\int_{0<|z|<1}\left[\int_{0}^{1}\left(\frac{\partial f}{\partial x}\!\left(X_{s}^{\varepsilon}(x)+\alpha
z,Y_{s}^{\varepsilon}(x,y)\right)\right)^{2}d\alpha\right]z^{2}h_{A}^{\varepsilon,1}(z)\,dz\,ds\\
&\quad\quad\,\,+2\int_{0}^{t}\int_{0<|z|<1}\left[\int_{0}^{1}\left(\frac{\partial f}{\partial y}\!\left(X_{s}^{\varepsilon}(x)+z,Y_{s}^{\varepsilon}(x,y)+\alpha u_{\beta}(z)\right)\right)^{2}d\alpha\right]u_{\beta}^{2}(z)h_{A}^{\varepsilon,1}(z)\,dz\,ds<\infty,
\end{align}
where for the first inequality we have used that
\begin{align*}
&\left(f\left(X_{s}^{\varepsilon}(x)+z,Y_{s}^{\varepsilon}(x,y)+u_{\beta}(z)\right)-f\left(X_{s}^{\varepsilon}(x),Y_{s}^{\varepsilon}(x,y)\right)\right)^{2}\\
&\quad =\left(\int_{0}^{1}z\frac{\partial f}{\partial x}\left(X_{s}^{\varepsilon}(x)+\alpha z,Y_{s}^{\varepsilon}(x,y)\right)d\alpha+\int_{0}^{1}u_{\beta}(z)\frac{\partial f}{\partial y}\left(X_{s}^{\varepsilon}(x)+z,Y_{s}^{\varepsilon}(x,y)+\alpha u_{\beta}(z)\right)d\alpha\right)^{2}\\
&\quad\leq 2z^{2}\int_{0}^{1}\left(\frac{\partial f}{\partial x}\left(X_{s}^{\varepsilon}(x)+\alpha z,Y_{s}^{\varepsilon}(x,y)\right)\right)^{2}d\alpha+2u_{\beta}^{2}(z)\int_{0}^{1}\left(\frac{\partial f}{\partial y}\left(X_{s}^{\varepsilon}(x)+z,Y_{s}^{\varepsilon}(x,y)+\alpha u_{\beta}(z)\right)\right)^{2}d\alpha.
\end{align*}
Then, \eqref{eq:Dynkin} follows immediately by taking expectation on both sides of \eqref{eq:ItoXYeps} and the change of variables $s=\alpha t$. Next, we show that $L_{\varepsilon}f$ is bounded on $\mathbb{R}^{2}$. Clearly, $\mathcal{D}_{\varepsilon}f$ is bounded on $\mathbb{R}^{2}$ by Assumption \ref{assump:UnifBdVol} and since $f\in C_{b}^{2}(\mathbb{R}^{2})$. Moreover,
\begin{align}
\left|\mathcal{I}_{\varepsilon}f(x,y)\right|&\leq\int_{A_{0}}\left(\int_{0}^{1}\left|\frac{\partial^{2}f}{\partial x^{2}}\left(x+\alpha
z,y+u_{\beta}(z)\right)\right|(1-\alpha)\,d\alpha\right)z^{2}c_{\varepsilon}(z)h(z)\,dz\\
&\quad\,+\int_{A_{0}}\left(\int_{0}^{1}\left|\frac{\partial^{2}f}{\partial y^{2}}\left(x,y+\alpha
u_{\beta}(z)\right)\right|(1-\alpha)\,d\alpha\right)u_{\beta}^{2}(z)c_{\varepsilon}(z)h(z)\,dz\\
&\quad\,+\int_{A_{0}}\left[\int_{0}^{1}\left(\left|\frac{\partial^{2}f}{\partial x\partial y}\left(x,y+\alpha
u_{\beta}(z)\right)\right|+\left|\frac{\partial^{2}f}{\partial x\partial y}\left(x+\alpha
z,y\right)\right|\right)d\alpha\right]zu_{\beta}(z)c_{\varepsilon}(z)h(z)\,dz<\infty,
\end{align}
which completes the proof of the lemma.\hfill $\Box$

\bigskip
\noindent
\textbf{Proof of Lemma \ref{lem:SmallTimeEstI1}.} Let us start by recalling the following trivial formula
\begin{align}\label{eq:ExpToTailProb}
\mathbb{E}\left((Z-K){\bf 1}_{\{Z>K\}}\right)=\int_{K}^{\infty}\mathbb{P}(Z>z)\,dz,
\end{align}
valid for any random variable $Z$ and any constant $K>0$. In particular, we can rewrite \eqref{eq:I1} as
\begin{align}
I_{1}=e^{x}\int_{K/e^{x}}^{\infty}\mathbb{P}\left(Y^{\varepsilon}_{t}(x,x)-x>\ln z\right)dz.
\end{align}
Assumption \ref{assump:UnifBdVol}, together with the fact that the jump sizes of $Y^{\varepsilon}(x,x)$ are bounded by $\beta_{\varepsilon}:=\ln(|\beta|(e^{\varepsilon}-1)+1)$, implies that
\begin{align}
V_{t}^{\varepsilon}:=\beta\int_{0}^{t}\sigma\left(X^{\varepsilon}_{s}(x)\right)d\widetilde{W}_{s}+\int_{0}^{t}\int_{\mathbb{R}_{0}}u_{\beta}(z)\,\widetilde{M}^{\varepsilon,1}_{A}(ds,dz),\quad
t\geq 0,
\end{align}
is a martingale, whose quadratic variation is such that $[V^{\varepsilon}]_{t}\leq\widetilde{D}\,t$ for any $t\geq 0$, for a constant $\widetilde{D}>0$, depending on $\varepsilon$, $\beta$ and $\|\sigma\|_{\infty}$. By equation (9) in~\cite{LepeltierMarchael:1976}, for any $t\geq
0$, $D>0$, and $\lambda>0$, we have
\begin{align}\label{Eq:CI0}
\mathbb{P}\left(\sup_{s\in[0,t]}\left|V_{s}^{\varepsilon}\right|\geq D\right)\leq 2\exp\left[-\lambda D+\frac{\lambda^{2}}{2}\widetilde{D}\,t\left(1+e^{\lambda\beta_{\varepsilon}}\right)\right].
\end{align}
Note that the above inequality holds trivially when $D<0$. By Assumption \ref{assump:UnifBdVol}, the drift $\gamma_{\varepsilon}$ of
$Y^{\varepsilon}(x,x)$ is bounded (but depends on $\varepsilon>0$, $|\beta|$, and $\|\sigma\|_{\infty}$) and, hence, for any $\lambda>1$ and any $t\in[0,1]$, we have
\begin{align}\label{eq:EstTailVeps}
I_{1}&\leq e^{x}\!\!\!\int_{K/e^{x}}^{\infty}\!\!\!\mathbb{P}\!\left(\!\sup_{s\in[0,t]}\!\left|V_{s}^{\varepsilon}\right|\!\geq\!\ln
z\!-\!t\left\|\gamma_{\varepsilon}\right\|_{\infty}\!\right)\!dz\!\leq\!2e^{x}\!\!\!\int_{K/e^{x}}^{\infty}\!\!\!\exp\!\left(\!-\lambda\!\left(\ln
z\!-\!t\left\|\gamma_{\varepsilon}\right\|_{\infty}\!\right)\!+\!\frac{\lambda^{2}}{2}\widetilde{D}\,t\left(1\!+\!e^{\lambda\beta_{\varepsilon}}\!\right)\!\right)\!dz\\
&=\frac{2K}{(\lambda-1)}\exp\left(\lambda\left\|\gamma_{\varepsilon}\right\|_{\infty}t+\frac{\lambda^{2}}{2}\widetilde{D}\,t\left(1+e^{\lambda\beta_{\varepsilon}}\right)\right)e^{-\lambda(\ln
K-x)}.
\end{align}
Therefore, for any $n\in\mathbb{N}$ and any $t\in[0,e^{-3\beta_{\varepsilon}}]$, by choosing any $\varepsilon\in(0,\ln((e^{(\ln K-x)/(2n)}-1)/|\beta|+1)\wedge 1)$ (so that $\beta_{\varepsilon}\leq(\ln K-x)/(2n)$) and $\lambda=-\ln t/(2\beta_{\varepsilon})$,
we obtain that
\begin{align}
I_{1}\leq 4K\exp\left(\frac{\left\|\gamma_{\varepsilon}\right\|_{\infty}}{2\beta_{\varepsilon}}(-t\ln
t)+\frac{\widetilde{D}\,t\,(\ln t)^{2}}{8\beta_{\varepsilon}^{2}}\left(1+t^{-1/2}\right)\right)t^{n}\leq 4K\exp\left(\frac{\left\|\gamma_{\varepsilon}\right\|_{\infty}}{2e\beta_{\varepsilon}}+\frac{4\widetilde{D}}{e^{2}\beta_{\varepsilon}^{2}}\right)t^{n},
\end{align}
which completes the proof of the lemma.\hfill $\Box$

\bigskip
\noindent
\textbf{Proof of Lemma \ref{lem:SmallTimeEstI3}.} We begin by introducing some additional notations. For any $\varepsilon>0$, let
\begin{align}\label{DfnLambda}
\Lambda_{\varepsilon}:=\int_{\{|z|>\varepsilon\}\cap A}e^{\left|u_{\beta}(z)\right|}h(z)\,dz<\infty,
\end{align}
which can be shown to be finite if either $\beta\leq-1$, or $\beta\geq 1$ and Assumption \ref{assump:SmthBdLevyDen}-(ii) holds true. Indeed, for $\beta\leq -1$, $A=\left(-\infty,\ln(1-\beta^{-1})\right)$ and, for $z<0$, $u_{\beta}(z)>0$, in which case, $e^{|u_{\beta}(z)|}=\beta(e^{z}-1)+1$ is clearly integrable on $\{z<-\varepsilon\}$. For $\beta\geq 1$, $A=\left(\ln(1-\beta^{-1}),\infty\right)$ and, for $z>0$, $u_{\beta}(z)>0$, in which case, again $e^{|u_{\beta}(z)|}=\beta(e^{z}-1)+1$ is integrable on $\{z>\varepsilon\}$ under Assumption \ref{assump:SmthBdLevyDen}-(ii). Next, for any $n\in\mathbb{N}$, and any collection of (fixed) times $0<s_{1}<\cdots<s_{n}$, let $Y^{\varepsilon}(x,y;\{s_{1},\ldots,s_{n}\}):=(Y_{t}^{\varepsilon}(x,y;\{s_{1},\ldots,s_{n}\}))_{t\geq 0}$ and $X^{\varepsilon}(x;\{s_{1},\ldots,s_{n}\}):=(X_{t}^{\varepsilon}(x;\{s_{1},\ldots,s_{n}\}))_{t\geq 0}$ be the solution to the following two-dimensional SDE
\begin{align}
Y_{t}^{\varepsilon}(x,y;\{s_{1},\ldots,s_{n}\})&=y+\int_{0}^{t}\gamma_{\varepsilon}\!\left(X_{s}^{\varepsilon}(x;\{s_{1},\ldots,s_{n}\})\right)ds+\beta\int_{0}^{t}\sigma\left(X_{s}^{\varepsilon}(x;\{s_{1},\ldots,s_{n}\})\right)d\widetilde{W}_{s}\\
&\quad\,+\int_{0}^{t}\int_{\mathbb{R}_{0}}\ln\left(\beta\left(e^{z}-1\right)+1\right)\widetilde{M}_{A}^{\varepsilon,1}(ds,dz)+\sum_{i:\,s_{i}\leq
t}\ln\left(\beta\left(e^{J_{\varepsilon}^{(i)}}-1\right)+1\right),& t &\geq 0,\\
X_{t}^{\varepsilon}(x;\{s_{1},\ldots,s_{n}\})&=x+\int_{0}^{t}\mu_{\varepsilon}\!\left(X_{s}^{\varepsilon}(x;\{s_{1},\ldots,s_{n}\})\right)ds+\int_{0}^{t}\sigma\left(X_{s}^{\varepsilon}(x;\{s_{1},\ldots,s_{n}\})\right)d\widetilde{W}_{s}\\
&\quad\,+\int_{0}^{t}\int_{\mathbb{R}_{0}}z\widetilde{M}_{A}^{\varepsilon,1}(ds,dz)+\sum_{i:\,s_{i}\leq t}J_{\varepsilon}^{(i)},& t &\geq 0.
\end{align}
It follows that, for any $t\geq 0$, we have
\begin{align}
\left(X_{s}^{\varepsilon}(x;\{s_{1},\ldots,s_{n}\}),Y_{s}^{\varepsilon}(x;\{s_{1},\ldots,s_{n}\})\right)_{s\in[0,t]}\!\ed\!\left(\!\left.\left(X_{s}^{o}(x),Y_{s}^{o}(x,y)\right)_{s\in[0,t]}\right|N_{t}^{\varepsilon}(A)\!=\!n,\tau_{1}\!=\!s_{1},\ldots,\tau_{n}\!=\!s_{n}\!\right).
\end{align}
Note that, given $N_{t}^{\varepsilon}(A)=n$, the times of jumps $\tau_{1},\ldots,\tau_{n}$ are distributed as the order statistics of $n$ independent
uniform $[0,t]$ random variables. Hence,
\begin{align}
\mathbb{P}\left(\left.\left|Y_{t}^{o}(x,x)-x\right|>\ln
z\,\right|N_{t}^{\varepsilon}(A)=n\right)=\frac{n!}{t^{n}}\idotsint\limits_{0<s_{1}<\cdots<s_{n}<t}\mathbb{P}\left(\left|Y_{t}^{\varepsilon}(x,x;\{s_{1},\ldots,s_{n}\})-x\right|>\ln
z\right)ds_{1}\cdots ds_{n},
\end{align}
and it is sufficient to find, for any $0<s_{1}<\cdots<s_{n}<t\leq 1$, a uniform bound on
\begin{align}
\frac{1}{n!}\int_{0}^{\infty}\mathbb{P}\left(\left|Y_{t}^{\varepsilon}(x,x;\{s_{1},\ldots,s_{n}\})-x\right|>\ln
z\right)dz=\frac{1}{n!}\mathbb{E}\left(e^{|Y_{t}^{\varepsilon}(x,x;\{s_{1},\ldots,s_{n}\})-x|}\right).
\end{align}
To do this, first note that
\begin{align}
\mathbb{E}\left(e^{|Y_{t}^{\varepsilon}(x,x;\{s_{1},\ldots,s_{n}\})-x|}\right)&=\mathbb{E}\left(\mathbb{E}\left(\left.e^{|Y_{t}^{\varepsilon}(x,x;\{s_{1},\ldots,s_{n}\})-x|}\right|\mathscr{F}_{s_{n}-}\right)\right)\\
&=\mathbb{E}\left(\left.\mathbb{E}\left(e^{|Y_{t-s_{n}}^{\varepsilon}(v+J_{\varepsilon},w)+u_{\beta}(J_{\varepsilon})-x|}\right)\right|_{(v,w)=(X_{s_{n}}^{\varepsilon}(x;\{s_{1},\ldots,s_{n-1}\}),Y_{s_{n}}^{\varepsilon}(x,x;\{s_{1},\ldots,s_{n-1}\}))}\right)\\
&\leq\mathbb{E}\!\left(\left.\!\mathbb{E}\!\left(e^{|Y_{t-s_{n}}^{\varepsilon}\!(v+J_{\varepsilon},w)-x|+|u_{\beta}(J_{\varepsilon})|}\right)\right|_{(v,w)=(X_{s_{n}}^{\varepsilon}\!(x;\{s_{1},\ldots,s_{n-1}\}),Y_{s_{n}}^{\varepsilon}\!(x,x;\{s_{1},\ldots,s_{n-1}\}))}\right),
\end{align}
where, again, $J_{\varepsilon}$ is a random variable having the density $g_{J}$, given as in \eqref{eq:DenJumpZ2epsA}, and is independent of $Y^{\varepsilon}(v,w)$. By \eqref{DfnLambda} and \eqref{eq:EstTailVeps}, for any $\lambda>1$ and $t\in[0,1]$, we have
\begin{align}
&\mathbb{E}\left(e^{|Y_{t-s_{n}}^{\varepsilon}(v+J_{\varepsilon},w)-x|+|u_{\beta}(J_{\varepsilon})|}\right){\leq}\int_{A}\left(e^{|w-x|}\,\mathbb{E}\left(e^{|Y_{t-s_{n}}^{\varepsilon}(v+{\theta},w)-w|}\right)\right)e^{|u_{\beta}(\theta)|}g_{J}(\theta;\varepsilon)\,d\theta\\
&\quad\,\leq e^{|w-x|}\left({\Lambda_{\varepsilon/2}}+\int_{A}\left(\int_{1}^{\infty}\mathbb{P}\left(\left|Y_{t-s_{n}}^{\varepsilon}(v+\theta,w)-w\right|>\ln
z\right)dz\right)e^{|u_{\beta}(\theta)|}g_{J}(\theta;\varepsilon)\,d\theta\right)\\
&\quad\,\leq e^{|w-x|}\left({\Lambda_{\varepsilon/2}}+\int_{A}\left(\int_{1}^{\infty}2\exp\left(-\lambda\left(\ln
z-t\|\gamma_{\varepsilon}\|_{\infty}\right)+\frac{\lambda^{2}}{2}\widetilde{D}\,\left(1+e^{\lambda\beta_{\varepsilon}}\right)\right)dz\right)e^{|u_{\beta}(\theta)|}g_{J}(\theta;\varepsilon)\,d\theta\right)\\
&\quad\,\leq 3\Lambda_{\varepsilon/2}\exp\left(\lambda\|\gamma_{\varepsilon}\|_{\infty}+\frac{\lambda^{2}}{2}\widetilde{D}\,\left(1+e^{\lambda\beta_{\varepsilon}}\right)\right)\frac{e^{|w-x|}}{\lambda-1},
\end{align}
where $\beta_{\varepsilon}=\ln(|\beta|(e^{\varepsilon}-1)+1)$ and $\widetilde{D}>0$ is a constant depending on $\varepsilon$, $\beta$ and
$\|\sigma\|_{\infty}$ as used in \eqref{Eq:CI0}. By choosing $\lambda=e^{1/\sqrt{n}}\leq 3$, we obtain that
\begin{align}
\frac{1}{n!}\mathbb{E}\left(e^{|Y_{t}^{\varepsilon}(x,x;\{s_{1},\ldots,s_{n}\})-x|}\right)&\leq\frac{3\Lambda_{\varepsilon/2}}{n!}\,\exp\left(3\|\gamma_{\varepsilon}\|_{\infty}+\frac{9}{2}\widetilde{D}\left(1+e^{3\beta_{\varepsilon}}\right)\right)\frac{\mathbb{E}\left(e^{|Y_{s_{n}}^{\varepsilon}(x,x;\{s_{1},\ldots,s_{n-1}\})-x|}\right)}{{e^{1/\sqrt{n}}-1}}.
\end{align}
Proceeding by induction, we conclude that
\begin{align}\label{InEDH0}
\frac{1}{n!}\mathbb{E}\left(e^{|Y_{t}^{\varepsilon}(x,x;\{s_{1},\ldots,s_{n}\})-x|}\right)\leq\frac{1}{n!}{\widehat{D}^{n+1}}\left(\frac{1}{e^{1/\sqrt{n}}-1}\right)^{n+1}.
\end{align}
where $\widehat{D}:=3\Lambda_{\varepsilon/2}\exp\left(3\|\gamma_{\varepsilon}\|_{\infty}+(9/2)\widetilde{D}\!\left(1+e^{3\beta_{\varepsilon}}\right)\right)$. Finally, by noting that
\begin{align}
(n+1)\ln\widehat{D}-(n+1)\ln\left(e^{1/\sqrt{n}}-1\right)-\ln n!=n\ln\widehat{D}+\frac{1}{2}n\ln n-n\ln n+n+O(\ln n)\rightarrow -\infty,\;\text{as }\,n\rightarrow\infty,
\end{align}
we conclude that the right-hand side of the inequality \eqref{InEDH0} converges to $0$.\hfill $\Box$

\bigskip
\noindent
\textbf{Proof of Theorem \ref{thm:SmallTimeIVLETF}.} We will only present the proof of the OTM case $K>e^{x}$, while the ITM case can be proved similarly. Let us recall the standard Black-Scholes formula,
\begin{align}
C_{\text{BS}}(t;x,K,\sigma)=e^{x}N\left(\frac{x-\ln K}{\sigma\sqrt{t}}+\frac{\sigma\sqrt{t}}{2}\right)-KN\left(\frac{x-\ln K}{\sigma\sqrt{t}}-\frac{\sigma\sqrt{t}}{2}\right),
\end{align}
and that, by definition, the implied volatility $\hat{\sigma}(t)$ is such that $C_{\text{BS}}(t;x,K,\hat{\sigma}(t))=\Pi(t;x,K,\beta)$. It is then clear that since $\Pi(t;x,K,\beta)\sim b_{1}t$ converges to $0$, as $t\to{}0$, we must have that $\lim_{t\rightarrow 0}\hat{\sigma}(t)\sqrt{t}=0$, otherwise, if $\limsup_{t\rightarrow 0}\hat{\sigma}(t)\sqrt{t}=c\neq{}0$, then
$$
	\limsup_{t\to{}0}\,C_{\text{BS}}(t;x,K,\hat\sigma(t))=\left\{
	\begin{array}{ll}
	e^{x}N\left(\frac{x-\ln K}{c}+\frac{c}{2}\right)-KN\left(\frac{x-\ln K}{c}-\frac{c}{2}\right),&\text{if }c\in (0,\infty)\\
	e^{x}, &\text{if }c=+\infty.\end{array}\right.
$$
In both cases, we would have a contradiction. Next, note that, since $C_{\text{BS}}(t;x,K,\hat{\sigma}(t))=C_{\text{BS}}(t\hat{\sigma}^{2}(t);x,K,1)$ and $t\hat{\sigma}^{2}(t)\to{}0$, as $t\to{}0$, \eqref{eq:AsyExpBSOTMOpt} implies that, as $t\to{}0$,
\begin{align}\label{eq:AsyExpBSOTMOpt2}
C_{\text{BS}}(t;x,K,\hat{\sigma}(t))=\frac{K\hat{\sigma}^{3}(t)t^{3/2}}{\sqrt{2\pi}(\ln K-x)^{2}}\exp\left(-\frac{(\ln
K-x)^{2}}{2\hat{\sigma}^{2}(t)t}-\frac{\ln K-x}{2}\right)+O\left(\left(\hat{\sigma}^{2}(t)t\right)^{5/2}\right),
\end{align}
which, together with \eqref{eq:1stOrderAsyOTMCallOptLETF}, implies that
\begin{align}\label{eq:OTMOptLeadComp}
b_{1}t\sim\frac{K\,e^{-(\ln K-x)/2}}{\sqrt{2\pi}(\ln K-x)^{2}}\left(t\hat{\sigma}^{2}(t)\right)^{3/2}\exp\left(-\frac{(\ln
K-x)^{2}}{2t\hat{\sigma}^{2}(t)}\right),\quad t\rightarrow 0,
\end{align}
or, equivalently,
%Clearly, $\limsup_{t\rightarrow 0}t\hat{\sigma}^{2}(t)\geq 0$. Assume first that $\limsup_{t\rightarrow 0}t\hat{\sigma}^{2}(t)=c\in(0,\infty)$. Then $\limsup_{t\rightarrow 0}\hat{\sigma}(t)=\infty$. Thus, we have
%\begin{align}
%\limsup_{t\rightarrow 0}\frac{K\,e^{-(\ln K-x)/2}}{\sqrt{2\pi}(\ln K-x)^{2}}\left(t\hat{\sigma}^{2}(t)\right)^{3/2}\exp\left(-\frac{(\ln
%K-x)^{2}}{2t\hat{\sigma}^{2}(t)}\right)=\frac{K\,e^{-(\ln K-x)/2}}{\sqrt{2\pi}(\ln K-x)^{2}}c^{3/2}e^{-(\ln K-x)^{2}/(2c)}\neq 0.
%\end{align}
%Hence, the right-hand side of \eqref{eq:OTMOptLeadComp} does not converge to $0$ while the left-hand side does, which is clearly a contradiction.
%Assume next that $\limsup_{t\rightarrow 0}t\hat{\sigma}^{2}(t)=\infty$. Then the right-hand side of \eqref{eq:OTMOptLeadComp} explodes to
%$\infty$, which is again a contradiction. Thus, we must have $\lim_{t\rightarrow 0}t\hat{\sigma}^{2}(t)=0$ and, by \eqref{eq:OTMOptLeadComp}, we
%see that
\begin{align}
\lim_{t\rightarrow 0}\left(-\frac{(\ln K-x)^{2}}{2t\sigma^{2}(t)}+\frac{3}{2}\ln\left(t\hat{\sigma}^{2}(t)\right)-\ln t+\ln\left(\frac{K\,e^{-(\ln
K-x)/2}}{\sqrt{2\pi}(\ln K-x)^{2}}\right)-\ln b_{1}\right)=0.
\end{align}
Finally, since $\lim_{t\rightarrow 0}t\hat{\sigma}^{2}(t)\ln(t\hat{\sigma}^{2}(t))=0$, we obtain that
\begin{align}\label{eq:1stOrderAsyOTMIV}
\lim_{t\rightarrow 0}\left(-\frac{(\ln K-x)^{2}}{2}-t\sigma^{2}(t)\ln t\right)=0\quad\Rightarrow\quad\sigma^{2}(t)\sim -\frac{(\ln K-x)^{2}}{2t\ln
t}=:\sigma_{1}(t)=\sigma_{1}(t;x,K),\quad t\rightarrow 0.
\end{align}
The first-order approximation term for the implied volatility $\hat{\sigma}^{2}(t)$ does not depend on the leverage ratio $\beta$.

To derive the second-order approximation for $\hat{\sigma}^{2}(t)$, let
\begin{align}\label{eq:2ndRemOTMIV}
\tilde{\sigma}_{2}(t)=\tilde{\sigma}_{2}(t;x,K,\beta):=\frac{\hat{\sigma}^{2}(t)}{\sigma_{1}(t)}-1.
\end{align}
Clearly, we have $\tilde{\sigma}_{2}(t)\rightarrow 0$ as $t\rightarrow 0$. By \eqref{eq:1stOrderAsyOTMCallOptLETF}, \eqref{eq:AsyExpBSOTMOpt2} and
\eqref{eq:1stOrderAsyOTMIV}, for any $\delta>0$, there exists $t_{0}>0$, such that for any $t\in(0,t_{0})$,
\begin{align}
b_{1}t+o(t)\leq\Pi(t;x,K)&\leq\frac{Kt^{3/2}\hat{\sigma}^{3}(t)}{\sqrt{2\pi}(\ln K-x)^{2}}\exp\left(-\frac{(\ln
K-x)^{2}}{2t\hat{\sigma}^{2}(t)}-\frac{\ln K-x}{2}\right)\left(1+t\hat{\sigma}^{2}(t)\right)\\
&=\frac{Ke^{-(\ln K-x)/2}\,t^{3/2}\sigma_{1}^{3/2}(t)}{\sqrt{2\pi}(\ln K-x)^{2}}\exp\left(-\frac{(\ln
K-x)^{2}}{2t\sigma_{1}(t)\left(1+\tilde{\sigma}_{2}(t)\right)}\right)\left(1+\mathcal{E}(t)\right),
\end{align}
where $\mathcal{E}(t):=(1+\tilde{\sigma}_{2}(t))^{3/2}(1+t\hat{\sigma}^{2}(t))-1\rightarrow 0$ as $t\rightarrow 0$. Dividing both sides by $t=e^{-(\ln K-x)^{2}/(2t\sigma_{1}(t))}$, we obtain
\begin{align}
b_{1}+o(1)\leq\frac{Ke^{-(\ln K-x)/2}\,t^{3/2}\sigma_{1}^{3/2}(t)}{\sqrt{2\pi}(\ln K-x)^{2}}\exp\left(\frac{(\ln K-x)^{2}\tilde{\sigma}_{2}(t)}{2t\sigma_{1}(t)\left(1+\tilde{\sigma}_{2}(t)\right)}\right)\left(1+\mathcal{E}(t)\right).
\end{align}
Hence, by rearranging the above inequality, we have
\begin{align}
\frac{\tilde{\sigma}_{2}(t)}{1+\tilde{\sigma}_{2}(t)}&\geq\frac{2t\sigma_{1}(t)}{(\ln
K-x)^{2}}\ln\left(\frac{\sqrt{2\pi}\left[b_{1}+o(1)\right](\ln
K-x)^{2}e^{(\ln K-x)/2}}{Kt^{3/2}\sigma_{1}^{3/2}(t)\left(1+\mathcal{E}(t)\right)}\right)\\
&=\frac{2t\sigma_{1}(t)}{(\ln K-x)^{2}}\ln\left(\frac{\sqrt{2\pi}\,b_{1}(\ln K-x)^{2}e^{(\ln
K-x)/2}}{Kt^{3/2}\sigma_{1}^{3/2}(t)}\right)+\frac{2t\sigma_{1}(t)}{(\ln
K-x)^{2}}\ln\left(\frac{b_{1}+o(1)}{b_{1}\left(1+\mathcal{E}(t)\right)}\right)\\
&=\frac{1}{-\ln t}\ln\left(\frac{4\sqrt{\pi}\,b_{1}e^{(\ln K-x)/2}}{K\left|\ln K-x\right|}\left(\ln\left(\frac{1}{t}\right)\right)^{3/2}\right)+\frac{1}{-\ln
t}\ln\left(\frac{b_{1}+o(1)}{b_{1}\left(1+\mathcal{E}(t)\right)}\right)\\
&=:\sigma_{2}(t)+\widetilde{\mathcal{E}}(t)=\sigma_{2}(t;x,K,\beta)+\widetilde{\mathcal{E}}(t;x,K,\beta).
\end{align}
Since
\begin{align}
\sigma_{2}(t)=O\left(\frac{\ln\ln(1/t)}{\ln(1/t)}\right),\quad\widetilde{\mathcal{E}}(t)=o\left(\frac{1}{\ln(1/t)}\right),\quad\text{as
}\,t\rightarrow 0,
\end{align}
by solving $\tilde{\sigma}_{2}(t)$ from the above inequality, we obtain that
\begin{align}
\tilde{\sigma}_{2}(t)\geq\frac{\sigma_{2}(t)+\widetilde{\mathcal{E}}(t)}{1-\sigma_{2}(t)-\widetilde{\mathcal{E}}(t)}=\sigma_{2}(t)+\widetilde{\mathcal{E}}(t)+\frac{\left(\sigma_{2}(t)+\widetilde{\mathcal{E}}(t)\right)^{2}}{1-\sigma_{2}(t)-\widetilde{\mathcal{E}}(t)}=\sigma_{2}(t)+o\left(\frac{1}{\ln(1/t)}\right),\quad
t\rightarrow 0.
\end{align}
Proceeding similarly for the upper bound, we conclude that
\begin{align}\label{eq:2ndOrderAsyOTMIV}
\tilde{\sigma}_{t}(t)=\sigma_{2}(t)+o\left(\frac{1}{\ln(1/t)}\right),\quad t\rightarrow 0.
\end{align}
Combining \eqref{eq:1stOrderAsyOTMIV}, \eqref{eq:2ndRemOTMIV} and \eqref{eq:2ndOrderAsyOTMIV} completes the proof.\hfill $\Box$

%-----------------------------------------------------------------------------------
%
%       SECTION: 		Bibliography
%
%-----------------------------------------------------------------------------------

%\bibliographystyle{chicago}
%\bibliography{Bibtex-LETF}

\begin{thebibliography}{99}

\bibitem{Andersen2000}
L. Andersen and J. Andreasen.
\newblock {Jump diffusion models: Volatility smile fitting and numerical methods for pricing}.
\newblock {\em Review of Derivatives Research}, Vol. 4, pp. 231$-$262, 2000.

\bibitem{AhnHaughJain:2015}
A. Ahn, M. Haugh, and A. Jain.
\newblock {Consistent Pricing of Options on Leveraged ETFs}.
\newblock {\em SIAM Journal on Financial Mathematics}, Vol. 6, Issue 1, pp. 559$-$593, 2015.

\bibitem{Applebaum:2009}
D. Applebaum.
\newblock {\em L\'{e}vy Processes and Stochastic Calculus}, 2nd Edition.
\newblock {Cambridge Studies in Advanced Mathematics 116}, Cambridge University Press, New York, USA, 2009.

\bibitem{AvellanedaZhang:2010}
M. Avellaneda and S. Zhang.
\newblock {Path-Dependence of Leveraged ETF Returns}.
\newblock {\em SIAM Journal on Financial Mathematics}, Vol. 1, Issue 1, pp. 586$-$603, 2010.

\bibitem{CarrGemanMadanYor:2002}
P. Carr, H. Geman, D. B. Madan, and M. Yor.
\newblock {The Fine Structure of Asset Returns: An Empirical Investigation}.
\newblock {\em The Journal of Business}, Vol. 75, No. 2, pp. 305$-$332, 2002.

\bibitem{ChengMadhavan:2009}
M. Cheng and A. Madhavan.
\newblock {The Dynamics of Leveraged and Inverse-Exchange Traded Funds}.
\newblock {\em Journal of Investment Management}, Vol. 7, No. 4, pp. 43$-$62, 2009.

\bibitem{ContBentata:2015}
R. Cont and A. Bentata.
\newblock {Forward equations for option prices in semimartingale models}.
\newblock {\em Finance and Stochastics}, Vol. 19, No. 3, pp 617-65, 2015.

\bibitem{ContTankov:2004}
R. Cont and P. Tankov.
\newblock {\em Financial Modelling with Jump Processes}.
\newblock {Chapman \& Hall/CRC Financial Mathematics Series}, CRC Press, London, UK, 2004.

\bibitem{Dupire}
B. Dupire.
\newblock {Pricing with a Smile}.
\newblock {\em RISK Magazine}, Vol. 6, pp. 118-120, 1994.

\bibitem{FigueroaLopezForde:2012}
J. E. Figueroa-L\'{o}pez and M. Forde.
\newblock {The Small-Maturity Smile for Exponential L\'{e}vy Models}.
\newblock {\em SIAM Journal on Financial Mathematics}, Vol. 3, Issue 1, pp. 33$-$65, 2012.

\bibitem{FigueroaLopezGongHoudre:2012}
J. E. Figueroa-L\'{o}pez, R. Gong, and C. Houdr\'{e}.
\newblock {Small-Time Expansions of the Distributions, Densities, and Option Prices of Stochastic Volatility Models with L\'{e}vy Jumps}.
\newblock {\em Stochastic Processes and Their Applications}, Vol. 122, Issue 4, pp. 1808$-$1839, 2012.

\bibitem{FigueroaLopezLuoOuyang:2014}
J. E. Figueroa-L\'{o}pez, Y. Luo, and C. Ouyang.
\newblock {Small-Time Expansions for Local Jump-Diffusion Models with Infinite Jump Activity}.
\newblock {\em Bernoulli}, Vol. 20, No. 3, pp. 1165$-$1209, 2014.

\bibitem{FordeJacquierLee:2012}
M. Forde, A. Jacquier, and R. Lee.
\newblock {The Small-Time Smile and Term Structure for Implied Volatility under the Heston Model}.
\newblock {\em SIAM Journal on Financial Mathematics}, Vol. 3, Issue 1, pp. 690$-$708, 2012.

\bibitem{GaoLee:2014}
K. Gao and R. Lee.
\newblock {Asymptotics of Implied Volatility to Arbitrary Order},
\newblock {\em Finance and Stochastics}, Vol. 18, Issue 2, pp. 349$-$392, 2014.

\bibitem{GatheralHsuLaurenceOuyangWang:2012}
J. Gatheral, E. Hsu, P. Laurence, C. Ouyang, and T.-H. Wang.
\newblock {Asymptotics of Implied Volatility in Local Volatility Models}.
\newblock {\em Mathematical Finance}, Vol. 22, No. 4, pp. 591$-$620, 2014.

\bibitem{Kou:2002}
S. G. Kou.
\newblock {A Jump-Diffusion Model for Option Pricing}.
\newblock {\em Management Science}, Vol. 48, No. 8, pp. 1086$-$1101, 2002.

\bibitem{KouWang:2004}
S. G. Kou and H. Wang.
\newblock {Option Pricing under a Double Exponential Jump Diffusion Model}.
\newblock {\em Management Science}, Vol. 50, No. 9, pp. 1178$-$1192, 2004.

\bibitem{Leandre:1987}
R. L\'{e}andre.
\newblock {Densit\'{e} en Temps Petit d'un Processus de Sauts}.
\newblock {\em S\'{e}minaire de Probabilit\'{e}s XXI, Lecture Notes in Mathematics}, Vol. 1247, pp. 81$-$99, 1987.

\bibitem{LeeWang:2015}
R. Lee and R. Wang.
\newblock {How Leverage Shifts and Scales a Volatility Skew: Asymptotics for Continuous and Jump Dynamics}.
\newblock {\em Preprint}, 2015.

\bibitem{LepeltierMarchael:1976}
J. P. Lepeltier and B. Marchael.
\newblock {Probl\`{e}me des Martingales et \'{E}quations Diff\'{e}rentielles Stochastiques Associ\'{e}es \`{a} un Op\'{e}rateur Int\'{e}gro-Diff\'{e}rentiel}.
\newblock {\em Annales de I'Institut Henri Poincar\'{e}, Section B}, Tome 12, No. 1, pp. 43-103, 1976.

\bibitem{LeungLorigPascucci:2016}
T. Leung, M. Lorig, and A. Pascucci.
\newblock {Leveraged ETF Implied Volatilities from ETF Dynamics}.
\newblock {To appear in {\em Mathematical Finance}}, 2016.

\bibitem{LeungSantoli:2016}
T. Leung and M. Santoli.
\newblock {\em Leveraged Exchange-Traded Funds: Price Dynamics and Options Valuation}.
\newblock {Springer Briefs in Quantitative Finance}, Springer, 2016.

\bibitem{LeungSircar:2015}
T. Leung and R. Sircar.
\newblock {Implied Volatility of Leveraged ETF Options}.
\newblock {\em Applied Mathematical Finance}, Vol. 22, No. 2, pp. 162$-$188, 2015.

\bibitem{MadanCarrChang:1998}
D. B. Madan, P. P. Carr, and E. C. Chang.
\newblock {The Variance Gamma Process and Option Pricing}.
\newblock {\em European Finance Review}. Vol. 2, 79$-$105, 1998.

\bibitem{Marchal:2009}
P. Marchal.
\newblock {Small Time Expansions for Transition Probabilities of Some L\'{e}vy Processes}.
\newblock {\em Electronic Communications in Probability}, Vol. 14, pp. 132$-$142, 2009.

\bibitem{OksendalSulem:2005}
B. {\O}ksendal and A. Sulem.
\newblock {Applied Stochastic Control of Jump Diffusions}.
\newblock {Universitext}, Springer-Verlag, Berlin, Germany, 2005.

\bibitem{Zhu:2007}
G. D. Zhu.
\newblock {\em Pricing Options on Trading Strategies}.
\newblock {Ph.D Thesis}, New York University, 2007.

\end{thebibliography}

%-----------------------------------------------------------------------------------
%
%       SECTION: 		Figures
%
%-----------------------------------------------------------------------------------

\clearpage

\clearpage

\end{document}